\documentclass{elsart}

\usepackage{amssymb}
\usepackage{amsmath}
\usepackage{latexsym}
\usepackage{graphicx}

\begin{document}
\begin{frontmatter}

\title{A Modified Higher Order Godunov's Scheme for Stiff Source Conservative Hydrodynamics}

\author[a]{Francesco Miniati} \and \author[b]{Phillip Colella}
\address[a]{Physics Department, Wolfgang-Pauli-Strasse 16, ETH Z\"urich, CH-8093 Z\"urich}
\address[b]{Lawrence Berkeley National Laboratory, 1 Cyclotron Rd, Berkeley, CA 94720, USA}

\begin{abstract}
  We present an efficient second order accurate scheme to
  treat stiff source terms within the framework of higher order
  Godunov's methods. We employ Duhamel's formula to devise a modified
  predictor step which accounts for the effects of stiff source terms
  on the conservative fluxes and recovers the correct isothermal
  behavior in the limit of an infinite cooling/reaction rate. Source
  term effects on the conservative quantities are fully accounted for
  by means of a one-step, second order accurate semi-implicit
  corrector scheme based on the deferred correction method of Dutt et.
  al. We demonstrate the accurate, stable and convergent results of the
  proposed method through a set of benchmark problems for a variety of
  stiffness conditions and source types.
\end{abstract}
\begin{keyword}
\PACS 
\end{keyword}
\end{frontmatter}

\section{Introduction}

We wish to solve the following system of partial differential equations
describing a hydrodynamic flow with a stiff (energy) source term
\begin{equation} \label{hypsys:eq}
\frac{\partial U}{\partial t} + \sum _{d=1}^{D} \frac{\partial F_d(U)}{\partial x_d}  = S(U)
\end{equation}
where $D$ is the dimensionality of the problem, $U,~F(U),~S(U)$ are the
conservative variables, the conservative fluxes and the source term
respectively, given by
\begin{equation}
U=
\begin{pmatrix}
\rho \\
\rho u_1 \\
\vdots \\
\rho u_D \\
\rho E 
\end{pmatrix}
; \quad
F_d(U)=
\begin{pmatrix}
\rho u_d \\
\rho u_1 u_d + p \,\delta_{1d} \\
\vdots \\
\rho u_D u_d + p \,\delta_{Dd} \\
(\rho E + p) u_d 
\end{pmatrix}; \quad
S(U)=
\begin{pmatrix}
0 \\
0 \\
\vdots \\
0 \\
\rho\Lambda(e,\rho) 
\end{pmatrix}.
\label{eq:ugs}
\end{equation}
In the above equations, $\rho$ is the density, $u_d$ the velocity in
the $d$ direction, $E=e+ \sum_{d=1}^Du_d^2/2$ is the total specific energy
with, $e$, the specific internal energy.  $\Lambda(e,\rho)$ is the
term describing the source of specific internal energy.

In the following we consider the case of a stiff source term
corresponding to an endothermic process, such as occurs in radiative
losses.  In addition, we restrict our analysis to source types that,
at least near equilibrium, behave as a relaxation law.  In the stiff
case, the characteristic relaxation time scale for $S$ may be much
smaller than the CFL time step for the hydrodynamic waves. For that
reason, we would like to use a semi-implicit method, treating the
stiff source term implicitly, while using an explicit method for the
hyperbolic terms. However, the classical analysis of such fast
endothermic processes shows that, in the limit as the relaxation time
goes to zero, the gas can be described by the compressible flow
equations with an isothermal equation of state
\cite{vincentiKruger65}. Pember \cite{pember93} showed that the use of
formally second-order accurate semi-implicit methods such as Strang splitting,
or a second-order Godunov predictor-corrector method could lead
to a substantial loss of accuracy, due to inconsistencies between the
the characteristic tracing step
without sources and the effective limiting isothermal behavior.
%
Such inconsistencies between the flux calculation 
with the limiting isothermal equation of state can lead to dramatic
errors particularly at sonic points. 
Pember proposed various approaches to the problem based on classical
relaxation theory.  Roe and Hittinger \cite{roehit01} also addressed
the issues raised here in relation to Godunov's method with stiff
relaxation.  
In their approach they split the equations based on a splitting of state
space into stiff and non-stiff subspaces of the linearized source term
to obtain in the stiff limit formulations similar to ours. However,
neither Pember nor Roe and Hittinger did present a complete method
that is second-order accurate in both the stiff and non-stiff limits,
nor did they discuss the extension to more than one
dimension.

The problem of hyperbolic system with stiff relaxation has also 
been considered by other authors in the past mostly for
one-dimensional systems and within the framework of Runge-Kutta based
methods of lines.  In particular Jin \cite{jin95} designed second order
Runge-Kutta type splitting methods with the correct asymptotic limit.
Jin and Levermore's \cite{jinlevermore96} developed a semidiscrete
high resolution method which, in order to ensure the correct
asymptotic behavior, employes a linear combination of the conservative
fluxes for the homogeneous (i.e. without the relaxation term) and
equilibrium system.  The fluxes are computed with a higher order
Godunov's method and the scheme allows for a rapid transition between
the stiff and nonstiff regimes.  As the authors point out, however,
the upwind property of the scheme is not strictly guaranteed for all
stiffness conditions.  Finally, Caflisch, Jin and Russo
\cite{caflisch97} developed a scheme for hyperbolic systems with
relaxation that is uniformly accurate for various ranges of stiffness
conditions (see also
Ref.~\cite{jinpareschitoscani98,pareschirusso05}).

The aim of this paper is to build a higher order Godunov's method that
preserves the properties of robustness and accuracy across a variety
of stiffness conditions thus avoiding the problems described in
\cite{pember93}.  In particular, in order to preserve higher order
accuracy, we aim for a semi-implicit method that corresponds to a 
standard second-order Godunov method
of the appropriate hyperbolic problem for the stiff or non-stiff limits.  
To this end, we use second-order accurate deferred
corrections method of a type presented in \cite{dugrro00}, to obtain a
semi-implicit corrector that is  a special case of the algorithms
described in \cite{minion03}, although any implicit L-stable
second-order one-step method would be acceptable. The main new idea in
our work is contained in our treatment of the predictor step for
computing the hyperbolic fluxes, based on the derivation of a local
effective dynamics using Duhamel's formula. This leads to an explicit
predictor step that corresponds to that for a conventional
second-order Godunov method for Eq.~(\ref{hypsys:eq}) in the limit
where the relaxation time is comparable to or greater than the
hydrodynamic CFL time step; and to a second-order Godunov method for
the isothermal equations in the limit where the relaxation time is
much smaller than the hydrodynamic time step.  Our approach is
similar to that used in \cite{tcm05} for obtaining a well-behaved
numerical method for incompressible viscoelastic flows in both the
viscous and elastic limits; however, the details there are quite
different than those for the present setting.

The paper is organized as follows. In section~\ref{sdc:se} we
describe a second order accurate, semi-implicit corrector method based
on the deferred corrections ideas presented
in~\cite{dugrro00,minion03} to be used for the final source term
update.  In section~\ref{mgf:se}, based on Duhamel's formula, we work
out a modified formulation of Godunov's predictor step and flux
calculation suitable for the case of stiff source terms.  In
section~\ref{sa:se} we discuss stability issues for our approach, and
Section~\ref{lrnz:se} contains the extension of the method to the case
in which the source term depends both on the gas density as well as
the internal energy.  In section~\ref{test:se} we test the performance
of the code and demonstrate the accuracy of the method in various
stiffness conditions.  The paper concludes with section~\ref{concl:se}
where the main results of the paper are summarized.

\section{Semi-Implicit Predictor-Corrector} \label{sdc:se}
Our time-discretization for the source terms is a single-step,
second-order accurate scheme based on the deferred correction ideas in
Dutt, Greengard \& Rokhlin \cite{dugrro00}.  Given the system of
equations (\ref{hypsys:eq})
\begin{equation}\label{hypsyscom:eq}
\frac{\partial U}{\partial t}=-\nabla\cdot F+S(U)
\end{equation}
we aim for a scheme in which an explicit approach is retained for the
non-stiff conservative hydrodynamic term, $\nabla\cdot F$, and a
implicit method is employed for the stiff part of the equation, $S$.
The particular approach is a special case of a more general class of
semi-implicit methods by Minion ~\cite{minion03}.
Consider the first order system of ordinary differential
equations (ODEs)
\begin{equation} \label{sample:eq}
\frac{dY}{dt}=C(t,Y)
\end{equation}
\begin{equation}
Y(t=0)=Y_0
\end{equation}
with $Y\in\mathbb{R}^n$,
$C:\mathbb{R}\times\mathbb{R}^n\rightarrow\mathbb{R}^n$. 
In \cite{dugrro00}, Eq.~(\ref{sample:eq}) is reformulated in terms of
its equivalent Picard integral equation, to which a deferred
corrections algorithm is iteratively applied.  First an {\it error} is
estimated according to
\begin{equation} \label{eq:errdgr}
\tilde{\epsilon} (t) = Y_0 + \int^{t}_0 C\left[\tau,\tilde{Y}(\tau)\right] d\tau-\tilde{Y}(t) 
\hbox{ , } 0 \leq t \leq \Delta t
\end{equation}
where, $\tilde{Y}(t)$, is an initial guess to the solution
to be corrected iteratively.
Then a correction is computed by solving the error equation for the
correction $\delta(t)\equiv Y(t)-\tilde{Y}(t):$
\begin{gather}
\delta(t) = \int^{t}_0  \left\{ 
C \left[\tau,\tilde{Y}(\tau)+\delta(\tau)\right] - C \left[\tau,\tilde{Y}(\tau)\right] \right\} d\tau +\tilde{\epsilon}(t)
\label{eq:correq}
\\
{Y}(t) = \tilde{Y}(t) + \delta(t)
\hbox{ , } 0 \leq t \leq \Delta t.
\nonumber
\end{gather}

To complete the specification of the method, we need to choose a
quadrature scheme to replace the integrals in time by sums over a
finite number of points. The choice of quadrature method in the error
calculation (\ref{eq:errdgr}) and of the number of iterations
determines the accuracy of the method. However, as noted in
\cite{dugrro00}, the rate of convergence of the method is independent
of the accuracy of the quadrature rule used in the correction
calculation (\ref{eq:correq}). In particular, for stiff systems, one
uses a quadrature rule corresponding to backward Euler, replacing the
integrand by its linear approximation.  In the present case, we are
only interested in second-order accuracy, so we can use the
trapezoidal rule for the quadrature rule in the error calculation, and
iterate only once.

Our semi-implicit method will 
correspond to solving a collection of ODEs, one at each grid point
\begin{equation}
\frac{d U}{d t} = S(U) - (\nabla \cdot \vec{F})^{n + \frac{1}{2}}
\end{equation}
where we view the time-centered flux divergence as a constant source,
whose computation using a modified Godunov method is decried below.
Following \cite{minion03}, we solve the resulting collection of ODEs 
using the method described above. 
For our initial guess, we use
\begin{equation}
\tilde{U} =  U_0 + \left({\rm I}-\Delta t \nabla_U S |_{U_0}\right)^{-1} \left[ S(U_0) - (\nabla\cdot F)^{n+\frac{1}{2}}\right]\,\Delta t
\label{eq:wtld}
\end{equation}
where $U_0\equiv U(t_0)$.  In the above expression we have used
backward Euler to estimate the effects of the source term and we
have then Taylor expanded the implicit part of it into a linear form.
This yields a second order accurate estimate in the sense that:
$\tilde{U} - U(t_0+\Delta t)= O(\Delta t^2)$. Based on
Eq.~(\ref{eq:errdgr}) the error is then estimated as
\begin{equation}
\tilde{\epsilon}(\Delta t) = U_0+\frac{\Delta t}{2} \left[ S(\tilde{U})+S(U_0) \right]-
\Delta t \, (\nabla\cdot F)^{n+\frac{1}{2}} -\tilde{U}
\label{eq:err2}
\end{equation}
where we have used the trapezoidal rule to estimate the
integral of the source term.
The sought correction is obtained in implicit form by applying 
backward Euler to the integral in the correction equation
(\ref{eq:correq})
\begin{gather} \label{corr2:eq}
\delta(\Delta t)=({\rm I}-\Delta t\nabla_US|_{\tilde{U}})^{-1} \; \tilde{\epsilon}(\Delta t)
\\ \label{finalupdate:eq}
U(t_0+\Delta t)=\tilde{U}+\delta(\Delta t)
\end{gather}
From Eq.~(\ref{eq:err2})-(\ref{corr2:eq}) it is clear that the final
solution will have a truncation error $O(\Delta t^2)$ and global
second order accuracy in time.

Clearly in the non-stiff limit, as the contribution from the
term `$\Delta t\nabla_US|_{U}$' becomes negligible compared to those from
`${\rm I}$', the above scheme reduces to the usual second order accurate
explicit formulation
\begin{equation}
U(t_0+\Delta t) =  U_0 - \Delta t \, (\nabla\cdot F)^{n+\frac{1}{2}} +
\frac{\Delta t}{2} \left[ S(\tilde{U}) + S(U_0) \right].
\label{eq:nnstifflim}
\end{equation}

\section{Effective Dynamics and a Modified Godunov's Method} \label{mgf:se}

In order to compute the flux divergence $(\nabla \cdot
\vec{F})^{n+\frac{1}{2}}$, we use the quasilinear form of the
equations in primitive variables to extrapolate from cell centers to
cell faces
\begin{gather*}
\frac{\partial W}{\partial t} + \sum\limits_{d=1}^D A_d 
\frac{\partial W}{\partial x_d} = S^{(W)} (W) \\
S^{(W)} = \nabla_U W \, S(U).
\end{gather*}
In order to develop our formulation we will start
using $W = (\rho, \boldsymbol{u}, e)^T$, but will switch to the
usual set of primitive variables later in Sec.~\ref{modeig.sec}. 
Hereafter, we will
denote $S^{(W)} \equiv S$, dropping the superscript. We can also give
the evolution along the Lagrangian trajectories
\begin{gather*}
\frac{DW}{Dt} + \sum\limits_{d=1}^D A^L_d 
\frac{\partial W}{\partial x_d} = S(W) \\
A^L_d = A_d - u_d {\rm I} \hbox{ , } \frac{DW}{Dt} = 
\frac{\partial W}{\partial t} + (\boldsymbol{u} \cdot \nabla) W
\end{gather*}

We will derive from the quasilinear form of the equations 
a new system that includes, at least locally in time and state space,
the effects of the stiff source terms on the hyperbolic structure,
and use that quasilinear system to extrapolate from cell centers
to faces in a Godunov method. 

We first illustrate the approach for the case of a system of ODE. 
Consider the system of differential equations
\begin{gather}
\frac{dY}{dt}=BY+C(t),\quad  Y(t_0)=Y_0 \\
 Y:\mathbb{R}\rightarrow\mathbb{R}^n , \quad B\in\mathbb{R}^{n\times n}, 
\quad C:\mathbb{R}\rightarrow\mathbb{R}^n.
\end{gather} 
The evolution of the rate of change of $Y(t)$, namely $\delta Y\equiv
Y(t)-Y_0$, is then described by $d\delta Y/dt = B\delta Y+BY_0+ C(t)$
with $\delta Y(0)=0$.  According to Duhamel's formula,
\begin{equation} \label{duha1:eq}
\delta Y(t)  =  \int^t_{0}e^{(t-\tau)B} [BY_0+C(\tau)]\, d\tau .
\end{equation}
When the properties of $B$ lead to a stiff numerical problem, the
exponential term in the above integral is the one that changes most
rapidly, motivating the approximation
\begin{equation} \label{duha2:eq}
\delta Y(t) \approx  {\mathcal I}_B(\eta) \,[BY_0+C(0)] \,t 
\end{equation}
where
\begin{equation}  \label{iobt:eq}
{\mathcal I}_B(\eta) \equiv \eta^{-1} \, \int_0^\eta e^{\tau B} d\tau .
\end{equation}
\begin{equation} \label{eff:eq}
\frac{dY^{eff}}{dt}= {\mathcal I}_B(\eta)\, [BY_0+C(0)] .
\end{equation}
In what follows, we will take $\eta=O(\Delta t)$.
There are two distinguished limits to the effective equation.
First, if $||B\eta||\ll 1$, then
\begin{equation}
\mathcal{I}_B(\eta)=I+O(\eta)
\end{equation}
The second is when there is a single eigenmode of $B$ that is stiff relative
to the time scale defined by $\eta$.
Specifically, we assume that for some $v\in\mathbb{R}^n$
\begin{equation}
B=\tilde{B}-\lambda vv^T \quad \tilde{B}v=0,\thickspace v^T\tilde{B}=0
\end{equation}
with
\begin{equation}
\lambda\eta\gg 1,\quad ||\tilde{B}\eta||\ll 1.
\end{equation}

Here $\lambda^{-1}$ is the fast time scale that is stiff relative to $\eta$.
So we can write
\begin{equation}\label{ib1:eq}
\mathcal{I}_B(\eta)=(I-vv^T)+O\left(\eta,\frac{1}{\lambda}\right)
\end{equation}
and
\begin{equation}\label{ib2:eq}
\mathcal{I}_B(\eta)B=\tilde{B}+O\left(\eta,\frac{1}{\lambda}\right) .
\end{equation}
In this case, $\mathcal{I}_B(\eta)$ projects out the stiff dynamics, leaving only 
processes that are resolved on the $O(\eta)$ timescale.

We can use the effective equation (\ref{eff:eq}) with $\eta=\Delta t$ to compute a 
first-order accurate predictor step in a second-order accurate predictor-corrector.
\begin{equation}
Y^{\rm eff}(\Delta t) = Y(0)+\mathcal{I}(\Delta t)(BY(0)+C(0))\Delta t
\end{equation}
Then
\begin{equation}
\begin{split}
Y^{\rm eff}(\Delta t)-Y(\Delta t) & =O(\Delta t^2) ~~~~~~~~~~\text{ if (\ref{ib1:eq}) holds}
\\
& =\left(\Delta t^2+\frac{\Delta t}{\lambda}\right) ~~~\text{ if (\ref{ib2:eq}) holds}
\end{split}
\end{equation}

We apply this idea to the dynamics along Lagrangian trajectories.
We define
\begin{equation} \label{eq:dw0}
\delta W = W[\boldsymbol{x}(t),t] - W[\boldsymbol{x}(t_0),t_0] \equiv W-W_0
\end{equation}
%
%
%
and  
\begin{gather} \label{eq:w0}
\frac{D \delta W}{D t}+ G =S_0 +\dot{S}_0 \delta W  \\
G = \sum\limits^D_{d = 1} A^L_d \frac{\partial W}{\partial x_d}
\end{gather}
We have linearized the source term around the value of the state at
the beginning of the Lagrangian trajectory, with 
$\dot{S} = \nabla_W \cdot S$. 
%
By applying Duhamel's formula to Eq.~(\ref{eq:w0}) we obtain
\begin{equation} \label{dwle:eq}
\delta W(t) = \int_{t_0}^{t}
e^{(t-\tau)\dot{S}_0} (-G+S_0) d\tau .
\end{equation}
Following similar reasoning to the ODE case, we obtain
\begin{equation} \label{dsol:eq}
\frac{DW^{\rm eff}}{D t} +
\left(\sum _{d=1}^{D} {\mathcal I}_{\dot{S}_0}(\eta) \, A_d^L \, 
\frac{\partial W}{\partial x_d} \right) = 
{\mathcal I}_{\dot{S}_0}(\eta) \, S_0.
\end{equation}
\subsection{Characteristic Analysis} \label{charanalys:se}

We will use the quasilinear system~(\ref{dsol:eq}) with 
$ (\eta)= \Delta t/2$ to compute the Godunov predictor step.
In the non-stiff limit, this leads to an $O(\Delta t^2)$ error in the 
predicted values at cell faces which is sufficient for second-order
accuracy in the overall method.
In order to do that, we need to analyze the hyperbolic structure 
of those equations. 
Without loss of generality in the following subsections
we still consider the 1-dimensional case.
Also, in this section we will focus on the case
%
%
$\Lambda_\rho\equiv \partial \Lambda /\partial \rho =0, 
\Lambda_e \equiv \partial \Lambda /\partial e \neq 0$;
%
%
we will discuss the more general case in Sec. \ref{lrnz:se}.
With this choice of $\dot{S}_0$, from Eq.~(\ref{iobt:eq}) 
we obtain
\begin{equation}
{\mathcal I}_{\dot{S}_0}(\Delta t/2)=
\begin{pmatrix}
1 & 0 & 0 \\
0 & 1 & 0 \\
0 & 0 & \alpha
\end{pmatrix}
\end{equation}
where 
\begin{equation} \label{alpha:eq}
\alpha= \frac{e^{\frac{1}{2}\Lambda_e\Delta t }-1}{\frac{1}{2}\Lambda_e\Delta t} ~~~~~~~~~~0<\alpha<1.
\end{equation}
Thus, the presence of a stiff source term leads us to the 
transformations:
\begin{equation} \label{atransf:eq}
A\equiv A^L+u{\rm I}\rightarrow A^{\rm eff}= \begin{pmatrix}
0 & \rho & 0 \\
\frac{1}{\rho} \left(\frac{\partial p}{\partial\rho}\right)_e & 0 & \frac{1}{\rho} \left(\frac{\partial p}{\partial e}\right)_\rho \\
0 & \alpha \frac{p}{\rho} & 0
\end{pmatrix} +  u{\rm I}.
\end{equation}

\subsubsection{Modified eigenvalues} \label{modeig.sec}

Characteristic analysis of the matrix $A^{\rm eff}$
leads to the characteristic equation
\begin{equation}
\det (A^{\rm eff} - \lambda {\rm I}) = 
(\lambda-u)
\left[(\lambda-u)^2-\alpha\frac{p}{\rho^2}
\left(\frac{\partial p}{\partial e}\right)_\rho-
\left(\frac{\partial p}{\partial\rho}\right)_e\right]=0
\end{equation}
which admits the the familiar solutions
\begin{equation} \label{char:eq}
\lambda_0=u,\quad \lambda_\pm=u\pm\left[\alpha\frac{p}{\rho^2}\left(\frac{\partial p}{\partial e}\right)_\rho+\left(\frac{\partial p}{\partial\rho}\right)_e\right]^{\frac{1}{2}}.
\end{equation}
It appears from the above equation that 
the presence of the source term alters the sound speed according to
\begin{equation} \label{clim:eq}
c_s=\left( \frac{\partial p}{\partial \rho} \right)_s^\frac{1}{2} \rightarrow 
c_{\rm eff} = \left[ \alpha\frac{p}{\rho^2}\left(\frac{\partial p}{\partial e}\right)_\rho+\left(\frac{\partial p}{\partial\rho}\right)_e\right]^{\frac{1}{2}}.
\end{equation}
For a $\gamma$-law equation of state, we have
\begin{equation} \label{eos:eq}
p=(\gamma-1)\rho e
\end{equation}
\begin{equation} \label{ceff:eq}
c_{\rm eff} =\left\{\left[\alpha\left(\gamma-1\right)+1\right]\,\frac{p}{\rho}\right\}^{\frac{1}{2}}.
\end{equation}
Thus, in the limit of a negligible source term, $\alpha\rightarrow 1$, 
$c_{\rm eff}\rightarrow(\gamma p/\rho)^{1/2}$ and the polytropic behavior 
is recovered. However, in the limit of a stiff source term, 
$\alpha\rightarrow 0$, $c_{\rm eff}\rightarrow (p/\rho)^{1/2}$ 
and the isothermal regime is approached. This is also apparent from 
the expression for the rate of change of the internal energy along
Lagrangian trajectories
\begin{equation} \label{de:eq}
\frac{De}{Dt} =-\alpha\frac{p}{\rho}\frac{\partial u}{\partial x},
\end{equation}
suggesting the limit, $de\rightarrow0$ as $\alpha\rightarrow0$.
Notice that in our approach we retain the polytropic form of the
equation of state $p=(\gamma-1)\rho e$, $\gamma\neq 1$, but
we avoid differentiating it when the presence of source terms must
be taken into account.
Based on Eq.~(\ref{de:eq}) the pressure change is found to be 

\begin{equation} \label{dp:eq}
\frac{Dp}{Dt}=c^2_{\rm eff} \, \frac{D\rho}{Dt}= -\rho\,c^2_{\rm eff} \, \frac{\partial u}{\partial x}.
\end{equation}

Finally, we note that in general, in
$D-$dimensions, the above analysis applies unaltered to the linear 
operator, $A^{\rm eff}_{d}$, for each direction, $d$, 
after properly transforming 
$u\rightarrow u_d,~x\rightarrow x_d$.  In addition, $D-1$
equations are added describing the passive transport of momentum
components perpendicular to the $d$ direction, and the eigenvalue
$\lambda_0$ acquires multiplicity $D$.

\subsubsection{Modified Eigenvectors}

Given Eq.~(\ref{dp:eq}) we can now replace internal energy with
pressure and find out the expression for the eigenvectors for the
usual set of primitive variables. This reads

\begin{equation}
W=(\rho,u,p,s)^{\bf T},
\end{equation}
where in addition to density, velocity and pressure, we have also
included the specific entropy, $s=p\rho^{-\gamma}$ (useful, e.g., for
the case of hypersonic flows~\cite{minetal00}).
The change in specific entropy is given by
\begin{equation}
\frac{Ds}{Dt}=\rho^{-\gamma}\,\left(\frac{Dp}{Dt}-c^2\,\frac{D\rho}{Dt}\right)=
-\rho^{1-\gamma}\,\left(c^2_{\rm eff}-c^2\right)\,\frac{\partial u}{\partial x}
\equiv -\rho^{1-\gamma}\,\delta_{c^2} \,\frac{\partial u}{\partial x}
\end{equation}

The linear operator is

\begin{equation}
A^{\rm eff}=\begin{pmatrix}
0 & \rho & 0  & 0 \\
0 & 0 & \rho^{-1} & 0 \\
0 & \rho c^2_{\rm eff} & 0 & 0 \\
0 & \delta_{c^2}\rho^{1-\gamma} & 0 & 0 
\end{pmatrix} + u {\rm I} .
\end{equation}
The extra variable `$s$' results in an additional eigenvalue, 
$\lambda=u$, for the operator $A^{\rm eff}$.
The set of left and right eigenvectors are given respectively by
\begin{gather}
l_1 = \left(0, -\frac{\rho}{2c_{\rm eff}}, \frac{1}{2c^2_{\rm eff}}, 0 \right) \\
l_2 = \left(1, 0, -\frac{1}{c^2_{\rm eff}}, 0 \right) \\
l_3 = \left(0, 0, -\frac{\delta_{c^2}}{\rho^{\gamma}c^2_{\rm eff}}, 1\right) \\
l_4 = \left(0, \frac{\rho}{2c_{\rm eff}}, \frac{1}{2c^2_{\rm eff}}, 0\right)
\end{gather}

\begin{equation}
r_1=\begin{pmatrix}
1 \\
-\frac{c_{\rm eff}}{\rho} \\
c^2_{\rm eff} \\
\delta_{c^2}\rho^{-\gamma}
\end{pmatrix}; \quad
r_2=\begin{pmatrix}
1 \\
0 \\
0 \\
0
\end{pmatrix}; \quad
r_3=\begin{pmatrix}
0 \\
0 \\
0 \\
1
\end{pmatrix}; \quad
r_4=\begin{pmatrix}
1 \\
\frac{c_{\rm eff}}{\rho} \\
c^2_{\rm eff} \\
\delta_{c^2}\rho^{-\gamma}
\end{pmatrix}.
\end{equation}

\subsection{Godunov Predictor in One Dimension}

With the operator $A^{\rm eff}$ and the sets of left and right
eigenvectors that we have worked out in the previous section, the
Godunov predictor step is carried out as usual as follows.

First the local slopes are defined. In particular at each point left
and right one-sided slopes as well as cell centered slopes are
evaluated and then a final choice on the local slope $\Delta W_{i}$
is defined by using van Leer limiter.
The upwind, time averaged left $(-)$ and right $(+)$ states at
cell interfaces due to fluxes in the normal direction, $d$, are then
reconstructed as:

\begin{equation}
W_{i,\pm}=W^n_i+\frac{1}{2}\left(I-\frac{\Delta t}{\Delta x}A^{\rm eff}_{i}\right) P_{\pm}(\Delta W_{i})
\end{equation}
where
\begin{equation}
P_{\pm}(W)=\sum_{\pm\lambda_k>0}(l_k\cdot W)\cdot r_k.
\end{equation}
The source term component is likewise accounted for as
\begin{equation} \label{wpm:eq}
W_{i,\pm,d}=W_{i,\pm,d}+\frac{\Delta t}{2}  {\mathcal I}_{\dot{S}_0}(\Delta t/2) S_0.
\end{equation}

The fluxes at the cell faces $F_{i + \frac{1}{2}}$ are
computed by solving the Riemann problem with left and right states
given by $\left(W_{i,+} , W_{i+1,-}\right)$ to obtain
$W^{n+\frac{1}{2}}_{i+\frac{1}{2}}$ and computing $F_{i+\frac{1}{2}} =
F\left(W^{n+\frac{1}{2}}_{i+\frac{1}{2}}\right)$.

To modify this procedure to account for the effective dynamics, we use
the characteristic analysis of the effective dynamics to perform each
of the three steps. The projection operator and any limiting in
characteristic variables is done using the eigenvectors and
eigenvalues for the effective dynamics derived in Sec.
\ref{charanalys:se}.  Typical approximate Riemann solvers use
weak-wave approximations to compute the jumps, which only require the
linearized jump relations provided by the characteristic analysis for
the effective dynamics.  For the case of a polytropic gas, one can use
more nonlinear approximate Riemann solvers, e.g. two shock
approximations, to compute the jump relations, treating $1 + \alpha
(\gamma - 1)$ as an effective polytropic $\gamma$. This is done for
the results presented here. Finally, any entropy fixes required to
eliminate rarefaction shocks require only the sound speed, for which
we again use $c_{\rm eff}$.

\subsection{Extension to More than One Dimension}

For directionally unsplit schemes in $D$ dimensions an
additional step is required in order to correct the time-averaged
left/right states at cell interfaces, $W_{i,\pm,d}$ in
Eq.~(\ref{wpm:eq}), for the effects of $D-1$ fluxes perpendicular to the
cell interface normal direction. Based on Eq.~(\ref{dsol:eq})
the effect of the stiff source term would be accounted for by
carrying out for each additional direction, $d$, a transformation
\begin{equation} \label{ad:eq}
A_{d}\rightarrow  {\mathcal I}_{\dot{S}_0}(\Delta t/2) 
A_{L,d}+u_d{\rm I} \equiv A^{\rm eff}_{d}
\end{equation}
analogous to that described in Eq.~(\ref{atransf:eq}).
In the method proposed by \cite{colella90,saltzman94} the corrections
due to transverse fluxes are computed according to a conservative
scheme. For example in two dimensions
\begin{gather} \label{consupdate:eq}
W_{{i,j},\pm,x} = W_{{i,j},\pm,x} - \frac{\Delta t}{2 \Delta y} \nabla_U W \;
\left( F^y_{{i,j+\frac{1}{2}}} - F^y_{{i,j-\frac{1}{2}}} \right)
\end{gather}
where the input $W_{{i,j},\pm,x}$ is computed using a one-dimensional
Godunov calculation as in the previous section, as are the fluxes
$F^y_{{i,j+\frac{1}{2}}}$.  The notation in Eq.
(\ref{consupdate:eq}) indicates that primitive variables are
converted into conservative variables which are then updated through
conservative fluxes and then converted back into primitive form.
Thus, if we indicate with $\Delta
F^y_{\rho E}$ the undivided flux difference in the $d$ direction for
the total energy, the above transformations imply the following
correction
\begin{gather} \nonumber
\Delta F^y_{\rho E} \rightarrow \Delta F^y_{\rho E} + (\alpha -1) \, 
\frac{1}{2}\left(p_{i,j+\frac{1}{2}} + p_{i,j-\frac{1}{2}}\right) \, \left(u_{y,i,j+\frac{1}{2}}  -u_{y,i,j-\frac{1}{2}} \right)
\end{gather}
This modification leads to a pressure change in accord to Eq.~(\ref{dp:eq}).
Similarly, the entropy flux difference is modified as
\begin{gather} \nonumber
\Delta F^y_{\rho s} \rightarrow \Delta F^y_{\rho s} +
(\alpha-1)\,(\gamma-1)\, \frac{1}{2}\left[\left(\rho s \right)_{i,j+\frac{1}{2}}
+ \left(\rho s\right)_{i,j-\frac{1}{2} } \right] \,
\left(u_{y,i,j+\frac{1}{2}}  -u_{y,i,j-\frac{1}{2}} \right) .
\end{gather}

\section{Stability Considerations} \label{sa:se}

The method outlined above satisfy a number of conditions required for
numerical stability.  It is easy to see from Eq.~(\ref{de:eq}) that,
as $d\Lambda/de \rightarrow -\infty$, the internal energy decays rapidly
to its equilibrium value, and thereafter remains
constant, at that value.  Inspection of the characteristic
analysis shows that, in this limit, no information is carried along
the entropy wave corresponding to the eigenvalue $\lambda_0$.  This
means that the system of equations~(\ref{hypsys:eq}) effectively
reduces to the equilibrium system in which the internal energy is
fixed at its equilibrium value.  In addition, Eq.~( \ref{char:eq}) and
(\ref{clim:eq}) indicate that the so called {\it subcharacteristic
  condition} for the characteristic speeds at equilibrium is always
satisfied.  That is
\begin{equation} \label{subchar:eq}
\lambda_- < \lambda^{\rm eff}_- < \lambda_0  < \lambda^{\rm eff}_+ < \lambda_+
\end{equation}
where $\lambda^{\rm eff}_{+,-}$ and $\lambda_{+,0,-}$ are the
equilibrium and frozen eigenvalues, respectively.  The above
condition, while being necessary for the stability of our
linearized system \cite{withman74}, also guarantees that the numerical
solution tends to the solution of the equilibrium equation as the
relaxation time tends to zero \cite{chleli94}.  Since
the structure of the equations and the numerical framework, including
the Riemann solver, remain basically unaltered with respect to classic
Godunov's schemes except for the modification of the sound speed, one
expects the usual stability analysis to apply. The latter implies the
familiar CFL condition on the time step
\begin{equation} \label{cfl:eq}
\max (| \lambda_* |) \frac{\Delta t}{\Delta x} \le 1 \hspace{1 cm} *=-,0,+.
\end{equation}

As for the step involving the source update, stability analysis for
deferred correction methods of the type adopted here
was carried out through numerical experiments by Dutt et
al.~\cite{dugrro00}. Minion~\cite{minion03} 
extends such considerations to the case of
semi-implicit schemes as the one adopted here.  While the stability
and convergence properties of such schemes have not been fully
elucidated analytically, the analysis of these authors suggest that
they are in general very satisfactory and competitive with commonly
employed modern integration schemes.  

Here we show that, provided that the CFL condition in Eq.~(\ref{cfl:eq})
is satisfied, our method is A-stable, in the sense further
specified below.  To demonstrate this we apply the method to the
following model problem~\cite{caflisch97}
\begin{gather} \label{example:eq}
\nonumber
\frac{dY(t)}{dt}= A  Y + B Y \\
Y(0)=1
\nonumber
\end{gather} 
where $Y:\mathbb{R}\rightarrow\mathbb{C}$ and $A,B\in \mathbb{C}$
and represent the non-stiff and stiff part of the equation,
respectively. Using the notation
\begin{equation} \label{}
\nonumber
Y^{n+1} = P(z_1,z_2) \, Y^n ,
\end{equation} 
where $P(z_1,z_2)$ is the operator corresponding to the proposed method,
$z_1= A\Delta t$, $z_2=B \Delta t$, the stability region 
of the method, $P$, is defined as the region
$S_P = \{ z_1,z_2\in \mathbb{C} : \lvert P(z_1,z_2)\rvert<1 \} $.
A method, $P$, is A-stable if $S_P$ includes the plane 
$\mathbb{C}_- \equiv \{ z\in \mathbb{C} : \mathbb{R} (z) <0 \}$.
Inspection of equations~(\ref{eq:wtld})-(\ref{finalupdate:eq})
and simple algebraic manipulation lead to the expression
\begin{equation} \label{}
P(z_1,z_2) = [1+P_h(z_1)] \, \frac{1-\dfrac{3}{2}z_2}{(1-z_2)^2} +\frac{\dfrac{z_2}{2}}{1-z_2}
\end{equation} 
where $P_h$ is the hydrodynamic operator given by Godunov's method.
Using the CFL conditions, which ensures
$\lvert P(z_1,0)\rvert = \lvert 1+P_h(z_1)\rvert <1$, we find:
\begin{equation} \label{}
\lvert  P(z_1,z_2)\rvert ^2 < 
\frac{\left\lvert1-z_2-\dfrac{z_2^2}{2}\right\rvert}{(1-z_2)^4} < 1~~\forall z_2 :\mathbb{R}(z_2)<0.
\end{equation} 

\section{Extension to the Case $\Lambda_\rho \neq 0$} \label{lrnz:se}

When the source term depends on both the internal energy and
the gas density, $\Lambda_\rho \neq 0$ and we obtain
\begin{equation}
{\mathcal I}_{\dot{S}_0}(\Delta t)=
\begin{pmatrix}
1 & 0 & 0 \\
0 & 1 & 0 \\
(\alpha-1)\,\frac{\Lambda_\rho}{\Lambda_e} & 0 & \alpha
\end{pmatrix}
\end{equation}
with $\alpha$ defined in Eq.~(\ref{alpha:eq}). As a result
\begin{equation} \label{atransf2:eq}
A^{\rm eff}= \begin{pmatrix}
0 & \rho & 0 \\
\frac{1}{\rho} \left(\frac{\partial p}{\partial\rho}\right)_e & 0 & \frac{1}{\rho} \left(\frac{\partial p}{\partial e}\right)_\rho \\
0 & (\alpha-1) \frac{\Lambda_\rho}{\Lambda_e} \rho +\alpha \frac{p}{\rho} & 0
\end{pmatrix} +  u{\rm I}
\end{equation}
and the sound speed is now given by 
\begin{equation} \label{cseff2:eq}
c_{\rm eff} = \left\{ \left[(\alpha-1)\frac{\Lambda_\rho}{\Lambda_e}\rho+ \alpha\frac{p}{\rho}\right]
\frac{1}{\rho}\,\left(\frac{\partial p}{\partial e}\right)_\rho+\left(\frac{\partial p}{\partial\rho}\right)_e\right\}^{\frac{1}{2}}.
\end{equation}
Since $\alpha<1$ the term in squared brackets can become negative and
the sound speed imaginary. This behavior is related to the fact that
when $\Lambda_\rho\neq 0$ the gas is prone to thermal instability so
that the scheme cannot be simply generalized without taking into
account the specific properties of the source term. In general
one cannot expect an implicit method to work properly
except in the case of a system with a stable solution. 
For a $\gamma-$law equation of state, Eq.~(\ref{eos:eq}), $c_{\rm
  eff}^2 >0$ requires
\begin{equation} \label{stab:eq}
\frac{e}{\rho} \, \frac{\Lambda_e}{\Lambda_\rho} > 
\frac{ 1- \alpha}{\alpha \, (\gamma-1)}
\end{equation}
which is reminiscent of the thermal stability criterion~\cite{field65}, in
which case the term on the right-hand-side is 1.  In both the stiff
limit and non-stiff limits the RHS in Eq.~(\ref{stab:eq}) is of order
$-\Lambda_e \, \Delta t \gg 1$, indicating the potential for
triggering thermal instability of `numerical nature'.
For example, consider a source of the form
$\Lambda(\rho,e) = \rho^n \, \tilde{\Lambda}(\rho,e)$, so that 
\begin{equation}  \label{dldr:eq}
\Lambda_\rho(\rho,e) = n\rho^{-1}\, \Lambda(\rho,e)+ \rho^n \,
\tilde{\Lambda}_\rho(\rho,e).
\end{equation}
In general the former term can take both positive and negative values.
So even though it vanishes at equilibrium, its effect is destabilizing
and should be resolved in time. Depending on the definition of
$\Lambda$, it is possible that $\tilde{\Lambda}_\rho \geq 0$. Only in
this case is the latter term stabilizing and should contribute to the
sound speed in Eq.~(\ref{cseff2:eq}).

So our approach is to decouple any destabilizing component of
$\Lambda_{\rho}$, which we indicate with $\Lambda_{\rho,<}$,
from the characteristic analysis and associate it
explicitly with the source term so that its effect does not enter
the sound speed. In this case one would have to add a term
\begin{equation}  \label{deextra:eq}
\Delta p = \rho \Delta e = 
-(\alpha -1) \, \frac{\Lambda_{\rho,<}}{\Lambda_e}\,\rho^2 \,
\left(\nabla\cdot{\bf u}\right) \, \frac{\Delta t}{2}
\end{equation}
to the pressure component of the right hand side of
Eq.~(\ref{wpm:eq}).  In order to preserve second order accuracy, one
would require that the above term is resolved in time, i.e. the time
step is sufficiently small that $\Delta e <e$.  
The restriction placed by Eq.~(\ref{wpm:eq})
depends on the shape of the source function. However, near
equilibrium it does not play any role
because by definition to zeroth order the source term is zero. 
In fact we find that in all test cases with a density dependent source
explored below, including the one in section~\ref{amrshock:sec}, the
condition~(\ref{wpm:eq}) never constrained the time step.

\section{Tests} \label{test:se}

In this section we test the
performance of the proposed method in terms of both accuracy and
robustness.  As for the accuracy we consider a set of one dimensional
problems for which the analytic solution is known. In particular, we
use the test problems in \cite{pember93} for an
isothermal rarefaction fan and an isothermal shock wave and consider a
flow with a stiff relaxation term in the limit in which the relaxation
time approaches zero.  We then consider the case of a sinusoidal
perturbation with wave-vector both parallel (1-D) and skew (2-D) with
respect to the x-axis, and prove the second order accuracy of the
scheme for a variety of stiffness conditions. This we show
both for the case in which the source term does or does 
not depend on density. As for the robustness of
the method, we turn to multidimensional problems involving strong
shocks and large spatial gradients.  In particular we consider the
interaction of a strong shock with a spherical cloud, again assuming
a variety of stiffness conditions. The aim of the tests is to prove
the code performance in the case of complex and computationally more
challenging calculations.

As for the source term, in the following we mostly present results based 
on a relaxation law of the form
\begin{equation} \label{tcc:eq}
\Lambda = - {\rm K} \, \rho^\zeta \, \left[e - e_0\, \left(\frac{\rho}{\rho_0}\right)^\eta\right]
\end{equation}
where $K$ is the heat transfer coefficient and the internal energy, 
$e$, is related to pressure and density by the equation of state
(\ref{eos:eq}), and $\zeta$ and $\eta$ are parameters. 
When $\zeta=\eta=0$ the relaxation law expressed by 
Eq.~(\ref{tcc:eq}) reduces to the case studied 
in~\cite{pember93} and in the limit $K\rightarrow \infty$ 
it enforces isothermality. 

We test the case of a density dependent 
source term, by setting either $\zeta$ or $\eta$, or
both parameters, to a non zero value. Only the latter case 
is reported here although in all cases we obtain consistent
results in terms of convergence and accuracy.
When $\eta\neq 0$, Eq.~(\ref{tcc:eq})
forces the system towards an equilibrium configuration 
described by polytropic-like equation of state in which
$e=e_0\,(\rho/\rho_0)^\eta$. Thus, when 
$\eta>0$,~Eq.~(\ref{cseff2:eq}) implies an effective 
adiabatic index that, as it should be, tends to $(1+\eta)$, 
as $\alpha \rightarrow 0$.
\subsection{Riemann Problems}
We first consider one-dimensional Riemann problems 
described by the following initial conditions
\begin{equation} \label{ic:eq}
(\rho , u,p)  [x,t=0] = 
   \left\{ \begin{array}{lll}  
   (\rho_l, u_l,  p_l) & \mbox{if} &  x\le 0.5\\ 
   (\rho_r,  u_r, p_r) &  \mbox{if} &  x>0.5
    \end{array} \right. 
\end{equation}
and with a source term described by Eq.~(\ref{tcc:eq}).
Following \cite{pember93} we adopt
\begin{gather} 
K = 10^8 \\
e_0= \frac{p_{l,r}}{\rho_{l,r}\,(\gamma-1)}=1 \\
\gamma = 1.4 \\
\Delta x =  2.5\times 10^{-3}.
\end{gather}
The stiff nature of the problem is apparent as K$^{-1} \ll \Delta x
/c_{\rm eff}$, i.e. the relaxation time is much shorter than the
hydrodynamic time scale.
\subsubsection{Isothermal Rarefaction}
\begin{figure} 
\begin{center}
\includegraphics[height=0.5\textheight, scale=1.0]{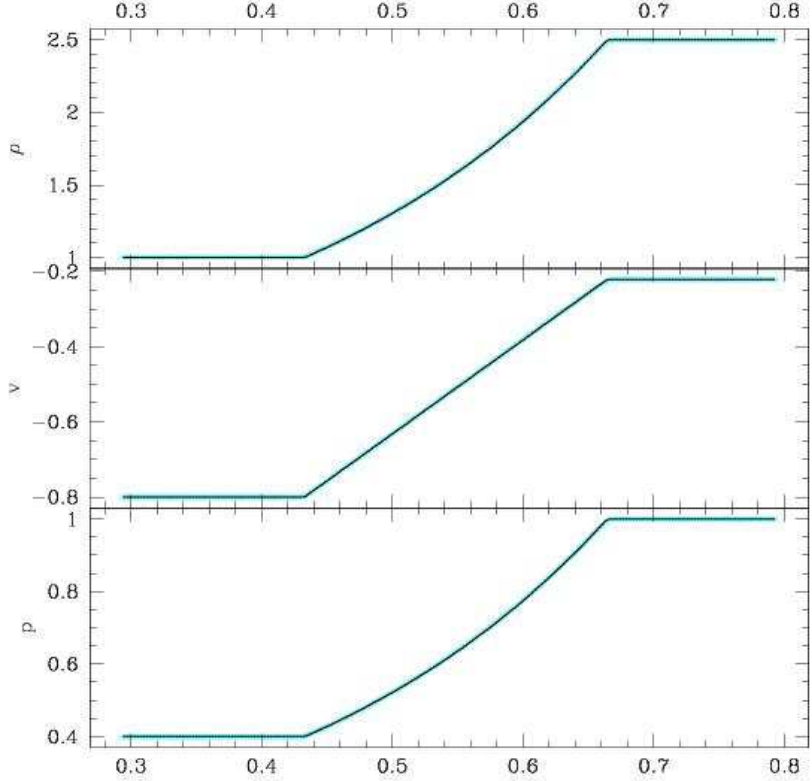}
\caption{Isothermal rarefaction wave. From top to bottom: density, 
velocity and pressure solutions, respectively. Open dots and 
solid line indicate the numerical and analytic solution, respectively.
The initial conditions are given in Eq.~(\ref{icfan:eq}) with 
$u_l=-0.8$. A mesh size $\Delta x  =  2.5\times 10^{-3}$ was employed.
\label{isofun:fig}}
\end{center}
\end{figure}
We begin by setting the state variables to the values
\begin{equation} \label{icfan:eq}
\begin{array}{lll}  
\rho_l = 1.0, & p_l  = 0.4, & u_l=-0.8 \\
\rho_r = 2.5, & p_r = 1.0, & u_r= u_l+0.5795
\end{array} 
\end{equation}
representing an isothermal rarefaction in the $\lambda_+$
characteristic family. For the calculation we employ a grid with
$N_{cell}=400$ grid cells~\cite{pember93}.  The results from the code
(open dots) are illustrated together with the analytic solution (solid
line) in Fig.~\ref{isofun:fig}. From top to bottom the plot shows the
density, velocity and pressure solutions at time $t=0.4$, respectively
(the same time as in~\cite{pember93}).  The solutions are free of
numerical artifact and well reproduce the analytic solution.  In
particular both the foot and the front edge of the rarefaction wave
are accurately reproduced as sharp features. In addition, there is no
numerical `kink' along the wave in correspondence of the sonic point,
that is the the eigenvalues $\lambda_+ = u+c_{\rm eff}$ changes
sign\footnote{ This occurs as the effective sound speed is $c_{\rm
    eff} \simeq 0.63$ and $u_l$ varies from $-0.8$ to $-0.2205$.}  as
it was noticed in the `non-stiff' schemes presented for comparison in
\cite{pember93}.  If we estimate the error in the numerical solution
as in~\cite{pember93}
\begin{equation}
\varepsilon = \frac{1}{N_{cell}} \sum_{i=1}^N \left| q_i^n - q_{iso}\left(x_{i+\frac{1}{2}},t^n \right) \right|
\end{equation}
that is the average of the absolute value of the difference between
the numerical and analytic result, we find that the error is
$\varepsilon_\rho= 4.2 \times 10^{-4}$ for the density and
$\varepsilon_u= 1.5 \times 10^{-4}$ for the velocity.  The latter is a
factor almost 20 smaller than obtained with the `frozen method'
proposed in~\cite{pember93}, most likely owing to the sharper
resolution of our method at the rarefaction front and foot.  This is
visible from comparing the analytic and numerical solutions in
Fig.~\ref{isofun:fig}. It is also consistent with the $L_\infty$ norm
of the errors (see Eq.~(\ref{lnorm_n:eq}) in Sec. \ref{convrate:se}),
which gives $\| \varepsilon_u \|_\infty = 7.3 \times 10^{-3}$ and $\|
\varepsilon_\rho \|_\infty = 1.6 \times 10^{-2}$, indicating a
localized error as opposed to one that is uniformly distributed.

\subsubsection{Shocks}
\begin{figure} 
\begin{center}
\includegraphics[width=0.50\textwidth,  scale=1.0]{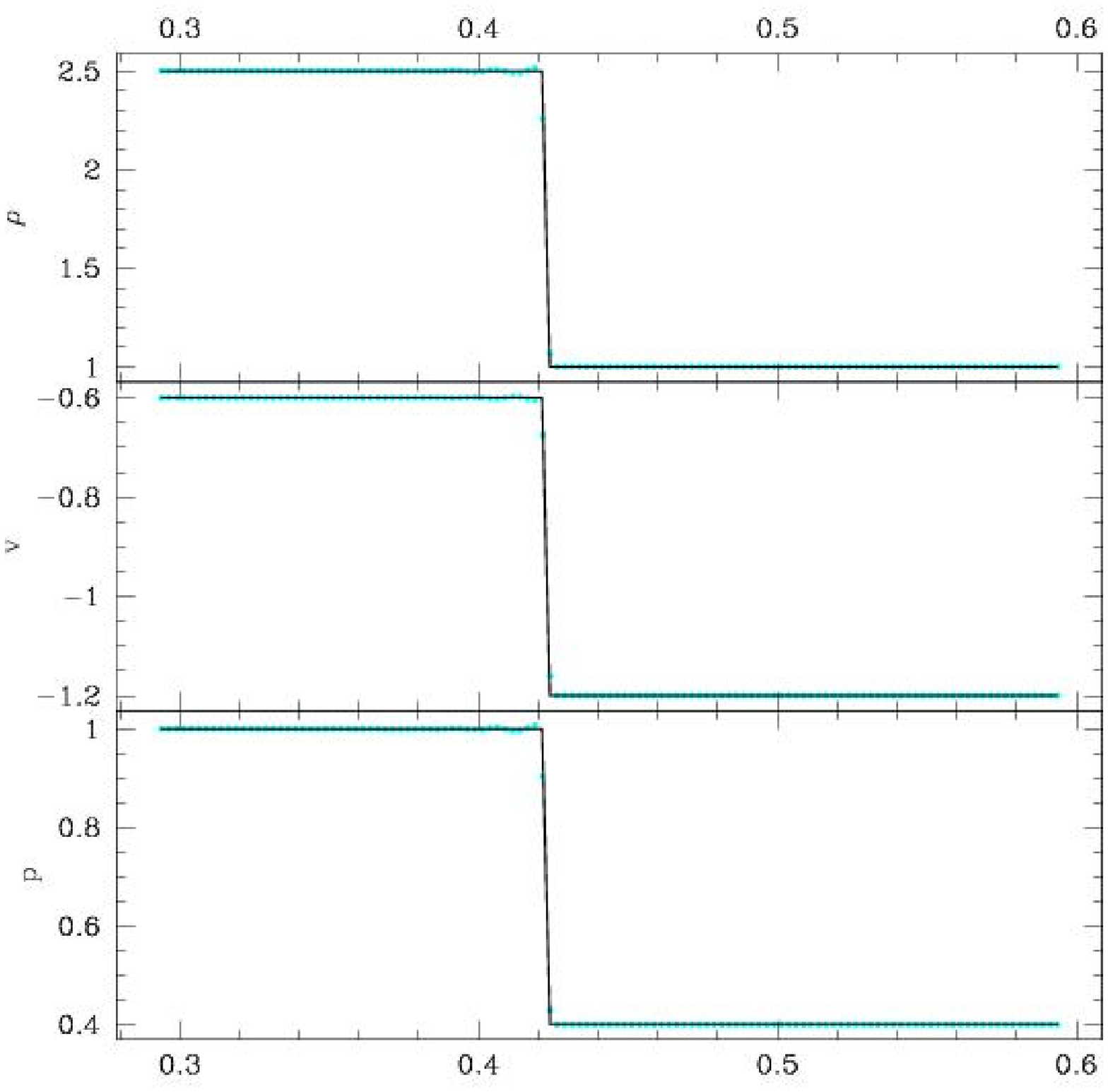}\includegraphics[width=0.50\textwidth, scale=1.0]{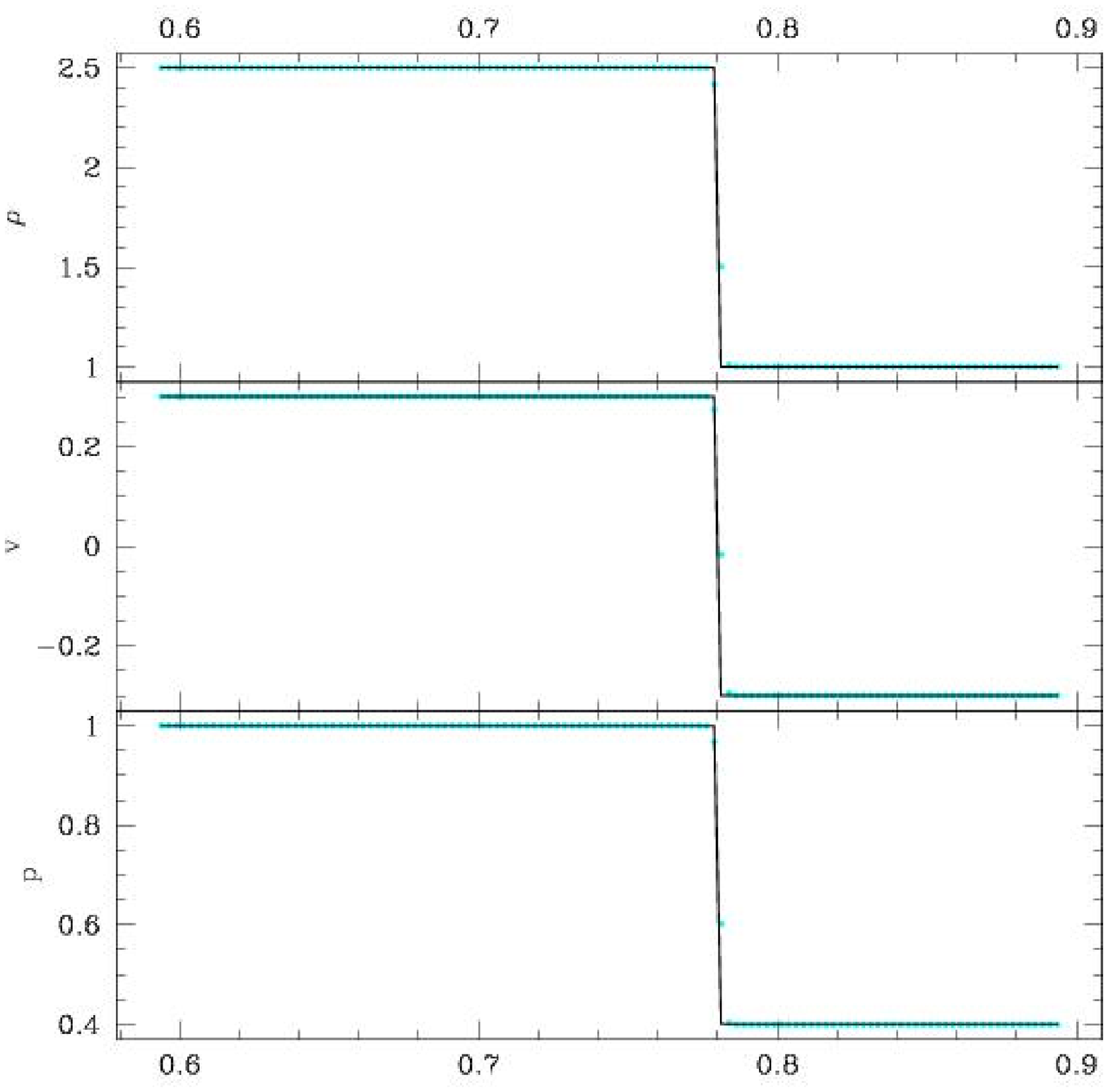}
\includegraphics[width=0.50\textwidth, scale=1.0]{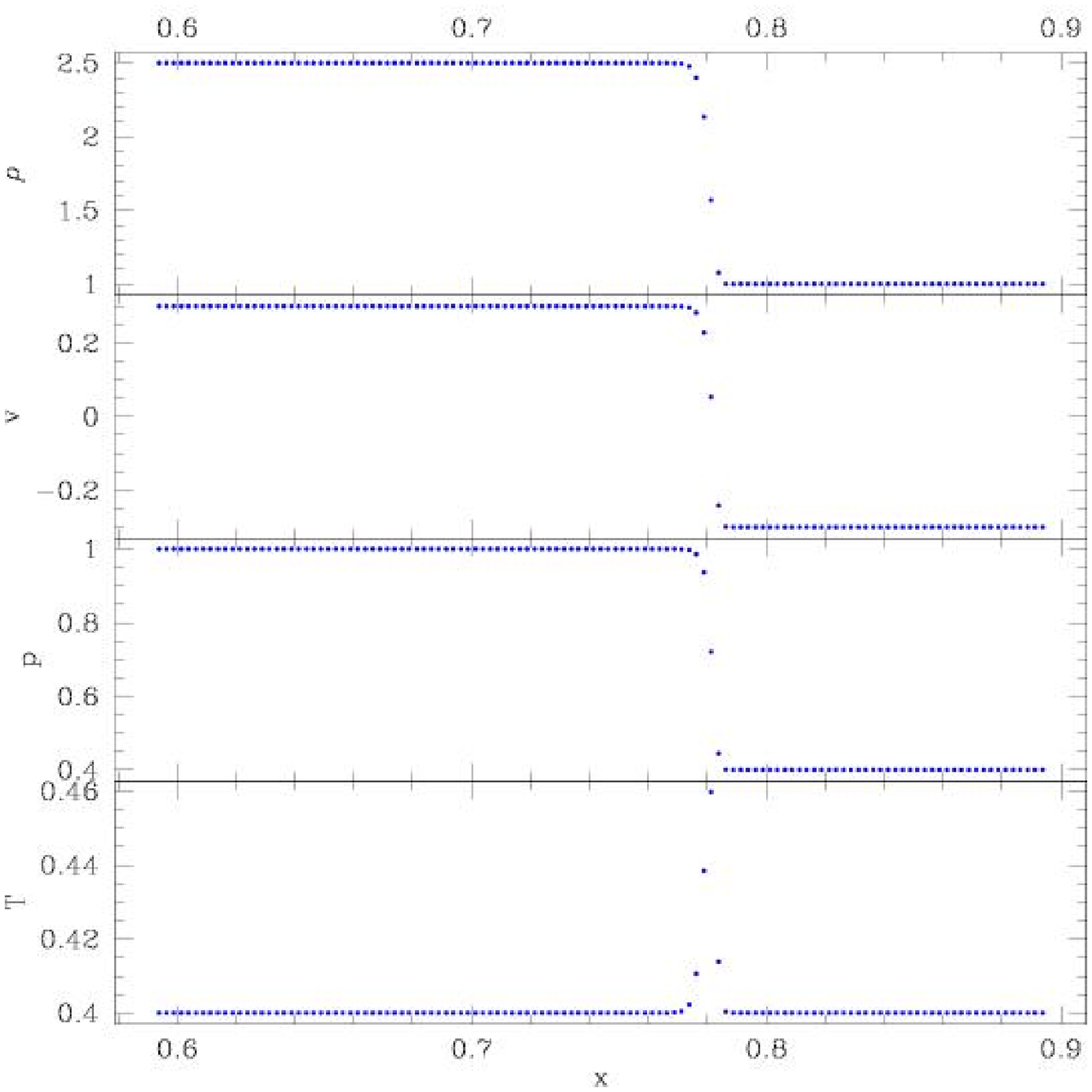}\includegraphics[width=0.50\textwidth,  scale=1.0]{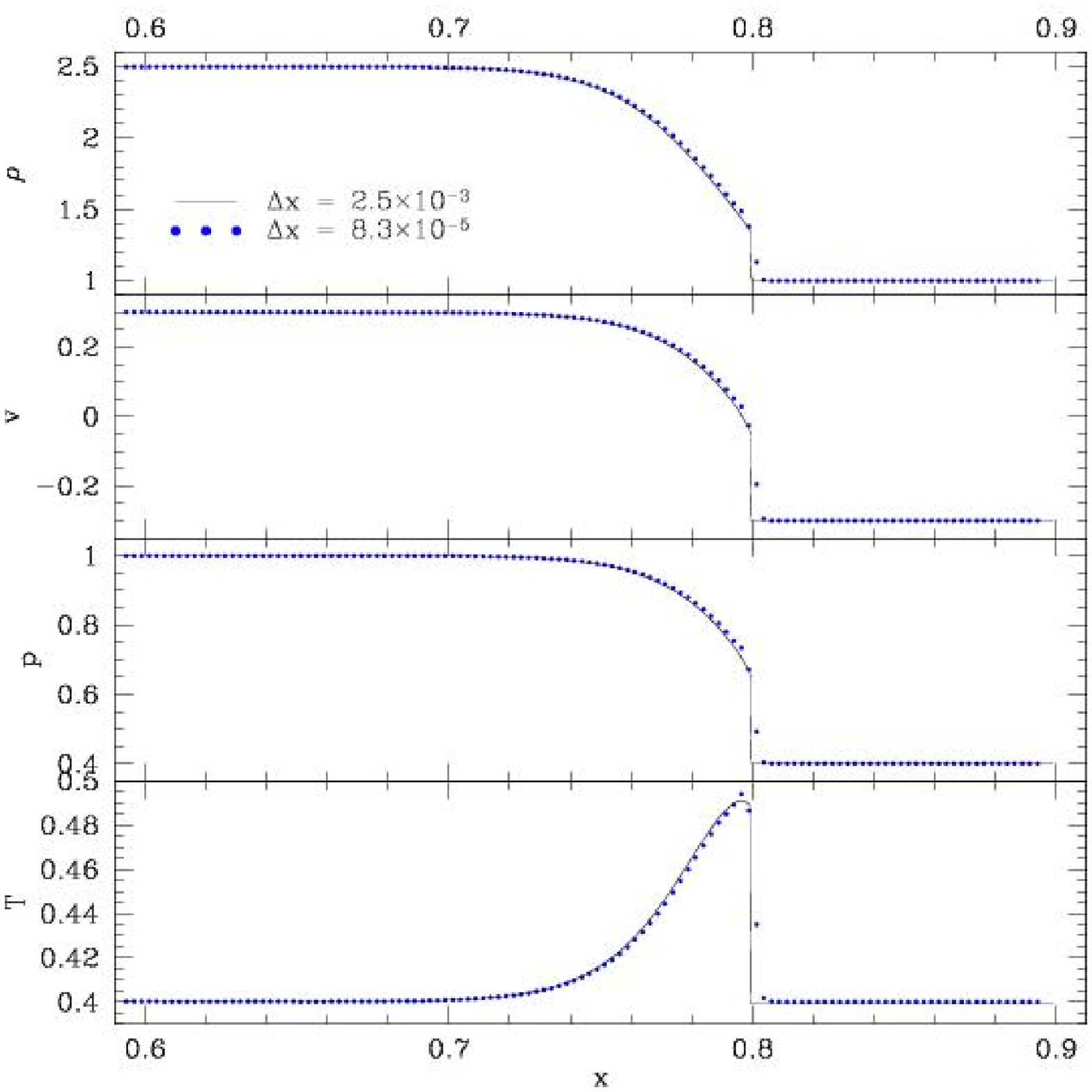}
\caption{Top panels: slow left-moving (left panel) 
and right-moving (right panel) isothermal shock waves.
In each panel, from top to bottom: density, 
velocity and pressure solutions, respectively. Filled dots and 
solid line indicate the numerical and analytic solution, respectively.
The initial conditions are given in Eq.~(\ref{icsho:eq}) with 
$u_r=-1.2$ (top left), and $u_r=-0.3$ (top right).
Bottom: slow right-moving quasi-isothermal shocks. In both cases we set
$u_r=-0.3$ and use a heat transfer coefficient $K=500$ (bottom left) 
and $K=50$ (bottom right) respectively.
The solid line in the bottom right panel corresponds to a solution 
obtained with a much higher resolution run using 
$\Delta x = 8.3\times 10^{-5}$ so that the reaction time is resolved.
In all other cases a mesh size $\Delta x  =  2.5\times 10^{-3}$ was employed.
\label{isoshock:fig}}
\end{center}
\end{figure}
Next we study the case of an isothermal shock with initial conditions
\begin{equation} \label{icsho:eq}
\begin{array}{lll}  
\rho_r = 1.0, & p_r = 0.4, & u_r= u_r \\
\rho_l = 2.5, & p_l  = 1.0, & u_l= u_r+0.6.
\end{array} 
\end{equation}
The numerical results (solid dots) are shown together with the
analytic solution (solid line) in Fig.~\ref{isoshock:fig} for two
different values of $u_r$, namely $-1.2$ (top left) $-0.3$ (top right)
producing isothermal shock fronts slowly moving to the left and the
right, respectively. Neither artificial viscosity nor flattening was
employed and a van Leer's limiter was used.  Overall the algorithm
performs very well.  The shock positions accord with the analytic
value.  The shocks are well captured within a couple of zones,
indicating that the properties of the scheme have not degraded with
respect to the non-stiff case.  We notice that minor oscillations
appear in a few zones downstream the left moving shock front.  These
have not been introduced by our method for treating the stiff source,
but rather, are due to the fact that the dissipation in a Godunov
method vanishes for slowly-left-moving shocks, such as the one being
computed here.  A thorough discussion on this is found
in~\cite{woodwardcolella84}.  In particular, we find that the same
oscillations appear in the purely hydrodynamic version of the
algorithm with the relaxation term turned off, if we use an adiabatic
index $0<\gamma-1 \ll 1$ in order to mimic isothermality.

Finally, in the bottom right panel the same initial conditions  
as in the top right panel are used but in combination 
with a much smaller heat transfer
coefficient, $K=500$ (bottom left) and  $K=50$  (bottom right). 
In these cases $K^{-1} \le \Delta x /c_{\rm eff}$ and
$K^{-1} \ge \Delta x /c_{\rm eff}$, respectively, 
so that while the gas behavior is not strictly isothermal,
the relaxation time is still relatively short. Assessing the algorithm
performance for this situation is of relevance as well, as in
general stiffness of the conditions will
vary across a flow.  The bottom panels
of Fig.~\ref{isoshock:fig}, in addition to density, velocity and
pressure also present results for the temperature.
As it appears from this plot, the numerical solution is very
satisfactory, without numerical artifacts or oscillations.
For the case $K=50$, we take the test one step further and compare
the result obtained using the current grid settings (solid dots), 
that is $N_{cell}=400,~\Delta x=2.5 \times 10^{-3}$,
with those from a much higher resolution run (solid line) in which 
$N_{cell}=12000,~\Delta x=8.33 \times 10^{-5}$ so that the reaction 
time is fully resolved.
This comparison nicely shows that our modified scheme is able to capture
the correct behavior of the flow even at intermediate stiffness conditions
away from a purely isothermal behavior.
\subsection{Convergence Rates in Smooth Flows} \label{convrate:se}
In this section we test the convergence of the method by studying 
the case of a smooth flow with the following initial conditions
\begin{gather}
\rho  =  \rho_0 +\frac{{A}}{2}\left[\cos(2\pi\, {\bf k \cdot r}) + 1 \right] \\
p     =  p_0   =  0.5 \\
u_x   =  u_{x0} = 0.3 \\ 
u_y = u_{y0} = 0.5
\end{gather}
where ${\bf r}$ is the position vector and we use $\rho_0  =\gamma = 1.4$.
The above initial conditions produce a sinusoidal wave with amplitude ${A}$
propagating in the domain along the direction defined by the vector
${\bf k}$.  While we have experimented with various values for the parameters 
${A}$, ${\bf k}$ and $K$, below we present results for a few cases only, 
summarized in Table~\ref{runset.tab}. 

\begin{table}  
\caption{Run Set for Convergence Study\label{runset.tab} with Relaxation Law Eq.~(\ref{tcc:eq})}
\begin{tabular*}{\textwidth}{@{\extracolsep{\fill}}lcccccc}
\hline \hline 
run & ${A}$ & ${\bf k}$ & $K$ & $\zeta$  & $\eta$ & Note \cr
\hline \cr
A  &  $10^{-2}$ & $(1,0)$ & $K=1$ & 0 & 0 &   \cr 
B &  $10^{-2}$ & $(1,0)$ & $K=50$   & 0 & 0 &   \cr 
C  &  $10^{-2}$ & $(1,0)$ & $K=10^8$    & 0 & 0 &   \cr 
D  &  $10^{-2}$ & $(2/\sqrt{5},1/\sqrt{5})$ & $K=1$   & 0 & 0 & \cr 
E &  $10^{-2}$ & $(2/\sqrt{5},1/\sqrt{5})$ & $K=50$  & 0 & 0 & \cr 
F  &  $10^{-2}$ & $(2/\sqrt{5},1/\sqrt{5})$ & $K=10^8$& 0 & 0 & \cr 
G  &  $10^{-2}$ & $(1,0)$ & $K=10^8$ & 0 & 0 &$\delta e/e_0=0.4$ \cr 
H  &  $10^{-2}$ & $(2/\sqrt{5},1/\sqrt{5})$ & $K=10^8$& 0 & 0 & $\delta e/e_0=0.4$  \cr 
I  &  $10^{-2}$ & $(1,0)$ & $K=10^8$ & 1 & 0.1 &  $\delta e/e_0=0.4$ \cr 
L  &  $10^{-2}$ & $(2/\sqrt{5},1/\sqrt{5})$  & $K=10^8$ & 1 & 0.1 & $\delta e/e_0=0.4$ \cr 
\hline
\end{tabular*}
\end{table}
\begin{table}  
\caption{Convergence Rates: 1-D Case: ${A}=10^{-2},~{\bf k}=(1,0)$\label{crit.tab}}
\begin{tabular*}{\textwidth}{@{\extracolsep{\fill}}lcccclcccc}
\hline 
\hline \cr
  &  \multicolumn{4}{c}{\bf density} && \multicolumn{4}{c}{\bf momentum} \cr
 \cline{2-5} \cline{7-10} \cr
$N_{\rm cell}$& $L_1$ & $L_2$ & $L_\infty$ & $R_1^\dagger$ && $L_1$ & $L_2$ & $L_\infty$ & $R_1^\dagger$ \cr 
\hline \cr
   &  \multicolumn{9}{c}{$K=1$}  \cr
\hline
32  &  4.3E-7 & 9.5E-7 & 2.8E-6 & --  && 6.3E-7 & 1.4E-7 & 4.0E-6 & -- \cr
64  &  1.1E-7 & 2.4E-7 & 7.0E-7 & 2.0 && 1.3E-7 & 2.8E-7 & 8.0E-7 & 2.3 \cr
128 &  2.7E-8 & 6.0E-8 & 1.8E-7 & 2.0 && 2.8E-8 & 6.3E-8 & 1.8E-7 & 2.2 \cr
256 &  6.8E-9 & 1.5E-8 & 4.4E-8 & 2.0 && 6.7E-9 & 1.5E-8 & 4.2E-8 & 2.1 \cr
\hline \cr
   &   \multicolumn{9}{c}{$K=50$}  \cr
\hline
32  &  4.0E-7 & 8.8E-7 & 2.6E-6 & --  && 7.8E-7 & 1.7E-6 & 4.9E-6 & -- \cr
64  &  1.1E-7 & 2.5E-7 & 7.2E-7 & 1.9 && 1.5E-7 & 3.2E-7 & 9.2E-7 & 2.4 \cr
128 &  3.1E-8 & 6.9E-8 & 2.0E-7 & 1.8 && 2.9E-8 & 6.5E-8 & 1.8E-8 & 2.4 \cr
256 &  8.5E-9 & 1.9E-8 & 5.4E-8 & 1.9 && 6.2E-9 & 1.4E-8 & 3.9E-8 & 2.2 \cr
\hline \cr
    &  \multicolumn{9}{c}{$K=10^8$}  \cr
\hline
32  &  4.1E-7 & 9.2E-7 & 2.7E-6 & --  && 8.2E-7 & 1.8E-6 & 5.91-6 & -- \cr
64  &  9.7E-8 & 2.1E-7 & 6.4E-7 & 2.1 && 1.6E-7 & 3.6E-7 & 1.0E-6 & 2.4 \cr
128 &  2.3E-8 & 5.2E-8 & 1.5E-7 & 2.1 && 3.6E-8 & 8.0E-8 & 2.3E-7 & 2.1 \cr
256 &  5.7E-9 & 1.3E-8 & 3.8E-8 & 2.0 && 8.5E-9 & 1.9E-8 & 5.4E-8 & 2.1 \cr
\hline
\hline
\end{tabular*}
\qquad\llap{$^\dagger$} $R_1$ is the convergence rate based on the $L_1$ errors.
\end{table}
In particular we consider a perturbation amplitude ${A}=10^{-2}$ and
both a wave-vector aligned with the grid ${\bf k}=(1,0)$ and skew with
respect to it, ${\bf k}=(2/\sqrt{5},1/\sqrt{5})$.  We adopt a source
term as given in Eq.~(\ref{tcc:eq}) with values of the transfer
coefficient $K=1, 50, 10^8$ to explore different regimes in which the
relaxation is resolved in time, is stiff as well as the intermediate
regime (cases A-F).  We then repeat case C and F but with the initial
value of the internal energy offset from the equilibrium value by
$\delta e/e_0 = 40 \%$ (cases G-H).  Finally, we consider the case in
which the source term depends both on density and internal energy, as
described by Eq.~(\ref{tcc:eq}).  In particular, we show results
concerning the case in which $K=10^8$, $\zeta=1$, $\eta=0.1$, $\delta
e/e_0 = 40 \%$ (cases I-L). Consistent test results were also found by
setting $\zeta=1$, $\eta=0$ as well as $\zeta=0$, $\eta=0.1$.

In order to measure the rate at which the numerical solution converges,
for each problem we carry out a set of 5 simulation runs employing 
$N_{cell}=32,64,128,256,512$ for a total range of 32.   Note
that the stiffness conditions do not change significantly as the grid
is refined within the range of resolutions considered here.  Also, the
smallness of the perturbations is such that the term given by
Eq.~(\ref{deextra:eq}), to be added to the energy in the predictor
step, is resolved in time
\begin{table}  
\caption{Convergence Rates: 2-D Case: ${A}=10^{-2},~{\bf k}=(2/\sqrt{5},1/\sqrt{5})$\label{crqit.tab}}
\begin{tabular*}{\textwidth}{@{\extracolsep{\fill}}lcccclcccc}
\hline 
\hline \cr
  &  \multicolumn{4}{c}{\bf density} && \multicolumn{4}{c}{\bf momentum} \cr
 \cline{2-5} \cline{7-10} \cr
$N_{\rm cell}$& $L_1$ & $L_2$ & $L_\infty$ & $R_1^\dagger$ && $L_1$ & $L_2$ & $L_\infty$ & $R_1^\dagger$ \cr 
\hline \cr
   &  \multicolumn{9}{c}{$K=1$}  \cr
\hline
32  &  3.2E-5 & 3.5E-5 & 5.3E-5 & --  && 4.3E-5 & 4.8E-5 & 7.0E-5 & -- \cr
64  &  8.6E-6 & 9.5E-6 & 1.4E-5 & 1.9 && 9.8E-6 & 1.1E-5 & 1.6E-5 & 2.1 \cr
128 &  2.2E-6 & 2.5E-6 & 3.6E-6 & 2.0 && 2.3E-6 & 2.6E-6 & 3.8E-6 & 2.1 \cr
256 &  5.7E-7 & 6.4E-7 & 9.2E-7 & 2.0 && 5.7E-7 & 6.3E-7 & 9.1E-7 & 2.0 \cr
\hline \cr
   &   \multicolumn{9}{c}{$K=50$}  \cr
\hline
32  &  6.2E-5 & 7.0E-5 & 1.0E-5 & --  && 3.6E-5 & 4.0E-5 & 5.9E-5 & -- \cr
64  &  1.2E-5 & 1.4E-5 & 2.0E-5 & 2.4 && 7.9E-6 & 8.8E-6 & 1.3E-5 & 2.4 \cr
128 &  2.7E-6 & 3.0E-6 & 4.3E-6 & 2.2 && 1.8E-6 & 1.9E-6 & 2.8E-6 & 2.1 \cr
256 &  6.2E-7 & 6.9E-7 & 1.0E-6 & 2.1 && 4.0E-7 & 4.5E-7 & 6.5E-6 & 2.2 \cr
\hline \cr
    &  \multicolumn{9}{c}{$K=10^8$}  \cr
\hline
32  &  7.4E-5 & 8.2E-5 & 1.2E-4 & --  && 4.5E-5 & 5.0E-5 & 7.3E-5 & -- \cr
64  &  1.6E-5 & 1.8E-5 & 2.6E-5 & 2.2 && 1.0E-5 & 1.1E-5 & 1.7E-5 & 2.2 \cr
128 &  3.8E-6 & 4.2E-6 & 6.1E-6 & 2.1 && 2.5E-6 & 2.8E-6 & 4.1E-6 & 2.0 \cr
256 &  9.4E-7 & 1.0E-6 & 1.5E-6 & 2.0 && 6.2E-7 & 6.9E-7 & 9.9E-6 & 2.0 \cr
\hline
\hline
\end{tabular*}
\qquad\llap{$^\dagger$} $R_1$ is the convergence rate based on the $L_1$ errors.
\end{table}
\begin{table}  
\caption{Convergence Rates: Off Equilibrium Case: $\delta e/e_0 =0.4$,~${A}=10^{-2}$,~$K=10^8$\label{crit2.tab}}
\begin{tabular*}{\textwidth}{@{\extracolsep{\fill}}lcccclcccc}
\hline
\hline  \cr
  &  \multicolumn{4}{c}{\bf density} && \multicolumn{4}{c}{\bf momentum} \cr 
 \cline{2-5} \cline{7-10} \cr
$N_{\rm cell}$& $L_1$ & $L_2$ & $L_\infty$ & $R_1^\dagger$ && $L_1$ & $L_2$ & $L_\infty$ & $R_1^\dagger$ \cr
\hline  \cr
   &  \multicolumn{9}{c}{${\bf k}=(1,0)$}  \cr
\hline
32  & 4.8E-7 & 1.1E-6 & 3.1E-6  &  --  && 7.6E-6 & 1.7E-6 & 4.8E-6  & -- \cr
64  & 1.0E-7 & 2.3E-7 & 6.9E-7  &  2.2 && 1.5E-7 & 3.3E-7 & 9.5E-6  & 2.3 \cr
128 & 2.4E-8 & 5.4E-8 & 1.6E-7  &  2.1 && 3.3E-8 & 7.3E-7 & 2.1E-7  & 2.2 \cr
256 & 5.9E-8 & 1.3E-8 & 3.9E-7  &  2.0 && 7.7E-9 & 1.7E-8 & 5.9E-8  & 2.1 \cr
\hline \cr
    &  \multicolumn{9}{c}{${\bf k}=(2/\sqrt{5},1/\sqrt{5})$}  \cr
\hline
32  &  7.4E-5 & 8.2E-5 & 1.2E-4  & --  && 4.4E-5 & 4.9E-5 & 7.2E-5 & --  \cr
64  &  1.6E-5 & 1.8E-6 & 1.7E-5  & 2.2 && 1.0E-5 & 1.1E-5 & 1.7E-5 & 2.1 \cr
128 &  3.8E-6 & 4.2E-6 & 6.2E-6  & 2.1 && 2.5E-6 & 2.8E-6 & 4.0E-6 & 2.0 \cr
256 &  9.4E-7 & 1.0E-6 & 1.5E-6  & 2.0 && 6.1E-7 & 6.8E-7 & 9.9E-7 & 2.0 \cr
\hline
\hline
\end{tabular*}
\qquad\llap{$^\dagger$} $R_1$ is the convergence rate based on the $L_1$ errors.
\end{table}
\begin{table}  
\caption{Convergence Rates: Off Equilibrium, $\rho$-dependent Source Case:  $\delta e/e_0 =0.4$,~$\zeta=1$,~$\eta=0.1$,~${A}=10^{-2}$,~$K=10^8$\label{crqit2.tab}}
\begin{tabular*}{\textwidth}{@{\extracolsep{\fill}}lcccclcccc}
\hline 
\hline \cr
  &  \multicolumn{4}{c}{\bf density} && \multicolumn{4}{c}{\bf momentum} \cr 
 \cline{2-5} \cline{7-10} \cr
$N_{\rm cell}$& $L_1$ & $L_2$ & $L_\infty$ & $R_1^\dagger$ && $L_1$ & $L_2$ & $L_\infty$ & $R_1^\dagger$ \cr 
\hline \cr
   &  \multicolumn{9}{c}{${\bf k}=(1,0)$}  \cr
\hline
32  &  2.6E-7 & 5.7E-7 & 1.8E-6 & --  && 9.0E-7 & 2.0E-6 & 5.7E-6 & -- \cr
64  &  6.1E-8 & 1.4E-7 & 4.2E-7 & 2.1 && 1.9E-7 & 4.1E-7 & 1.2E-6 & 2.2 \cr 
128 &  1.5E-8 & 3.3E-8 & 1.0E-7 & 2.0 && 4.2E-8 & 9.3E-8 & 2.7E-7 & 2.2 \cr
256 &  3.5E-9 & 7.7E-9 & 2.4E-8 & 2.1 && 1.0E-8 & 2.2E-8 & 6.5E-8 & 2.1 \cr
\hline \cr
    &  \multicolumn{9}{c}{${\bf k}=(2/\sqrt{5},1/\sqrt{5})$}  \cr
\hline
32  &  6.6E-5 & 7.4E-5 & 1.1E-4  & --  && 5.1E-5 & 5.7E-5 & 8.3E-5 & -- \cr
64  &  1.5E-5 & 1.7E-5 & 2.5E-5  & 2.1 && 1.2E-5 & 1.3E-5 & 1.9E-5 & 2.1 \cr 
128 &  3.7E-6 & 4.1E-6 & 5.9E-6  & 2.0 && 2.8E-6 & 3.1E-6 & 4.6E-6 & 2.1 \cr
256 &  9.1E-7 & 1.0E-6 & 1.5E-6  & 2.0 && 7.0E-7 & 7.7E-7 & 1.1E-6 & 2.0 \cr
\hline
\hline
\end{tabular*}
\qquad\llap{$^\dagger$} $R_1$ is the convergence rate based on the $L_1$ errors.
\end{table}

The convergence rate is measured using Richardson extrapolation.
Given the numerical result $q_{r}$ at a given resolution $r$ 
we first estimate the error at a given point $(i,j)$, as
\begin{equation} \label{numerr:eq}
\varepsilon_{r;i,j} = q_{r}(i,j) - \bar q_{r+1}(i,j)
\end{equation}
where $\bar q_{r+1}$ is the solution at the next finer resolution,
properly spatially averaged onto the coarser grid.
We then take the n-norm of the error
\begin{equation} \label{lnorm_n:eq}
L_n = \| \varepsilon_{r} \|_n =  \left( \sum |\varepsilon_{r;i,j} v_{i,j}|^n \right)^{1/n}
\end{equation}
where, $v_{i,j}=\Delta x^2$ is the cell volume, 
and estimate the convergence rate as
\begin{equation}
R_n = \frac{ \ln(L_n(\varepsilon_r)/L_n(\varepsilon_s)) }{ \ln (\Delta x_r / \Delta x_s) }.
\end{equation}
For each studied case listed Table~\ref{runset.tab}, we produce 
a corresponding table~(\ref{crit.tab}-\ref{crqit2.tab}) 
reporting the $L_1,~L_2$ and $L_\infty$ norms of the error 
as defined above. Inspection of
their values shows that the error drops with second order
accuracy, supporting our analysis in Sec.~\ref{sdc:se}..

A final experiment is designed to further prove that in the stiff
limit ($K=10^8$) the proposed scheme converges to the correct
asymptotic (isothermal) behavior. To do that we employ a Godunov 
method for the isothermal fluid equations and run
again case C in Table~\ref{runset.tab}.  A comparison of the solutions
obtained with the isothermal code and our proposed method is reported
in Table~\ref{asymlim.tab}.  It shows that the difference between the
two is always negligible with respect to the estimated truncation
error, thus validating our convergence study.

\begin{table}  
\caption{Comparison with a Purely Isothermal Solution: ${A}=10^{-2},~{\bf k}=(1,0)$\label{asymlim.tab}}
\begin{tabular*}{\textwidth}{@{\extracolsep{\fill}}lccclccc}
\hline 
\hline \cr
  &  \multicolumn{3}{c}{\bf density} && \multicolumn{3}{c}{\bf momentum} \cr 
 \cline{2-4} \cline{6-8} \cr
$N_{\rm cell}$& $L_1$ & $L_2$ & $L_\infty$ && $L_1$ & $L_2$ & $L_\infty$  \cr 
\hline \cr
32  &  2.7E-11 & 6.0E-11 & 1.7E-10  && 1.8E-10 & 4.0E-11 & 1.1E-10 \cr
128 &  2.7E-11 & 6.1E-11 & 1.7E-10  && 1.8E-10 & 4.0E-11 & 1.1E-10 \cr
256 &  2.7E-11 & 6.1E-11 & 1.7E-10  && 1.8E-10 & 4.0E-11 & 1.1E-10 \cr
512 &  2.7E-11 & 6.1E-11 & 1.7E-10  && 1.8E-10 & 4.0E-11 & 1.1E-10 \cr
\hline
\hline
\end{tabular*}
\end{table}
\subsection{Adaptive Mesh Refinement and Strong Shock Problems} \label{amrshock:sec}
In applications involving the interaction of strong shocks, 
it is useful to use the dissipation mechanisms described in 
\cite{colella90}, which generalize without modification to the present 
case. In addition, it is also desirable to couple this method to 
a block-structured adaptive mesh refinement (AMR)
\cite{bergercolella89,mico06a}.
In AMR calculations, the conservative 
variables are updated for the conservative fluxes in two steps.
The first step constitutes the main flux update and it simply consists in 
modifying the state variables $U$ for the total fluxes across the 
cell boundaries.
In addition, as part of the operations of synchronization among
different levels, the conservative variables at the coarse-fine
grid interfaces are further updated for the flux difference
between the level on which they are defined and the next finer level.
This operation is referred to as {\it refluxing} and it is 
aimed at preserving the conservative character of the numerical 
scheme when applied to a hierarchy of nested grids.

For the purpose of the current discussion, 
the effect of this operation can be expressed as 
\begin{equation} \label{reflux:eq}
U \rightarrow U - \frac{\Delta t}{\Delta x} \delta F
\end{equation}
where $\delta F$ is the difference between the fluxes at the 
coarse-fine interface computed on a given level and the next finer
level. In AMR calculations refluxing on a given level 
is enforced as a separate operation, after the source update
and the main flux update have been carried out
on that level and also on all finer levels.
Therefore, an additional measure must be taken to ensure 
that the effects of refluxing
are also subjected to the action of the deferred corrections
(just like the flux update does).
Thus, inspection of Eq.~(\ref{eq:wtld})-(\ref{corr2:eq})
indicates that the flux correction must be modified according to
\begin{equation} \label{refluxcorr:eq}
\delta F \rightarrow 
\left\{ 
\left({\rm I}-\Delta t \nabla_US|_{U_0}\right)^{-1} +
\left({\rm I}-\Delta t \nabla_US|_{\tilde{U}}\right)^{-1}
\left[{\rm I}-\left({\rm I}-\Delta t \nabla_US|_{U_0}\right)^{-1} \right] \right\} \, 
\delta F .
\end{equation}

In the following we employ an AMR code and carry out a calculation
involving the interaction of a spherical overdense region with a
strong hydrodynamic shock to assess the robustness of our proposed
numerical method.  We assume a cloud overdensity with respect to the
ambient medium $\chi=10$ and a shock Mach number ${\mathcal M}=10$.
\begin{figure} 
\begin{center}
\includegraphics[width=0.50\textwidth,  scale=1.0]{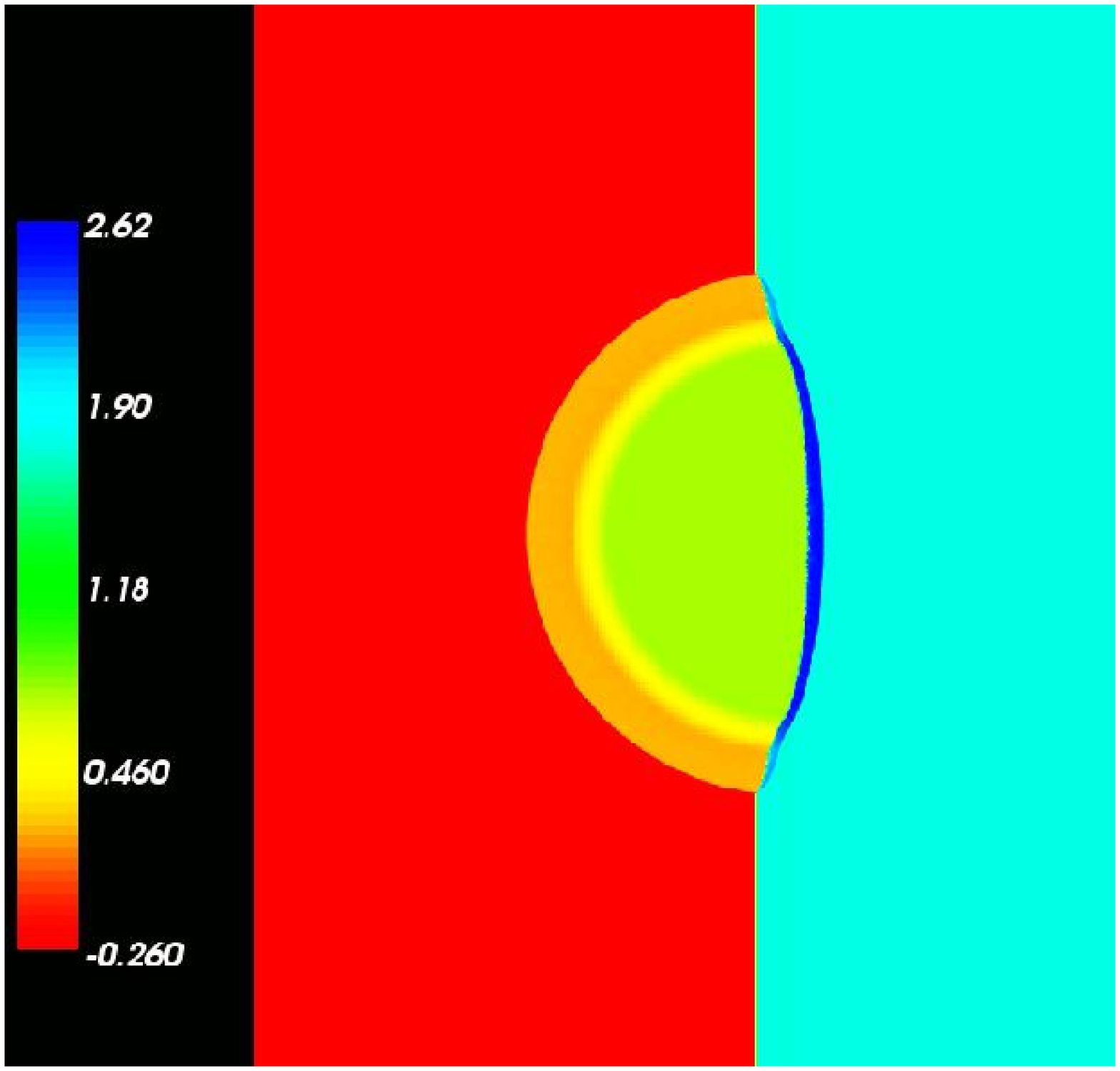}\includegraphics[width=0.50\textwidth, scale=1.0]{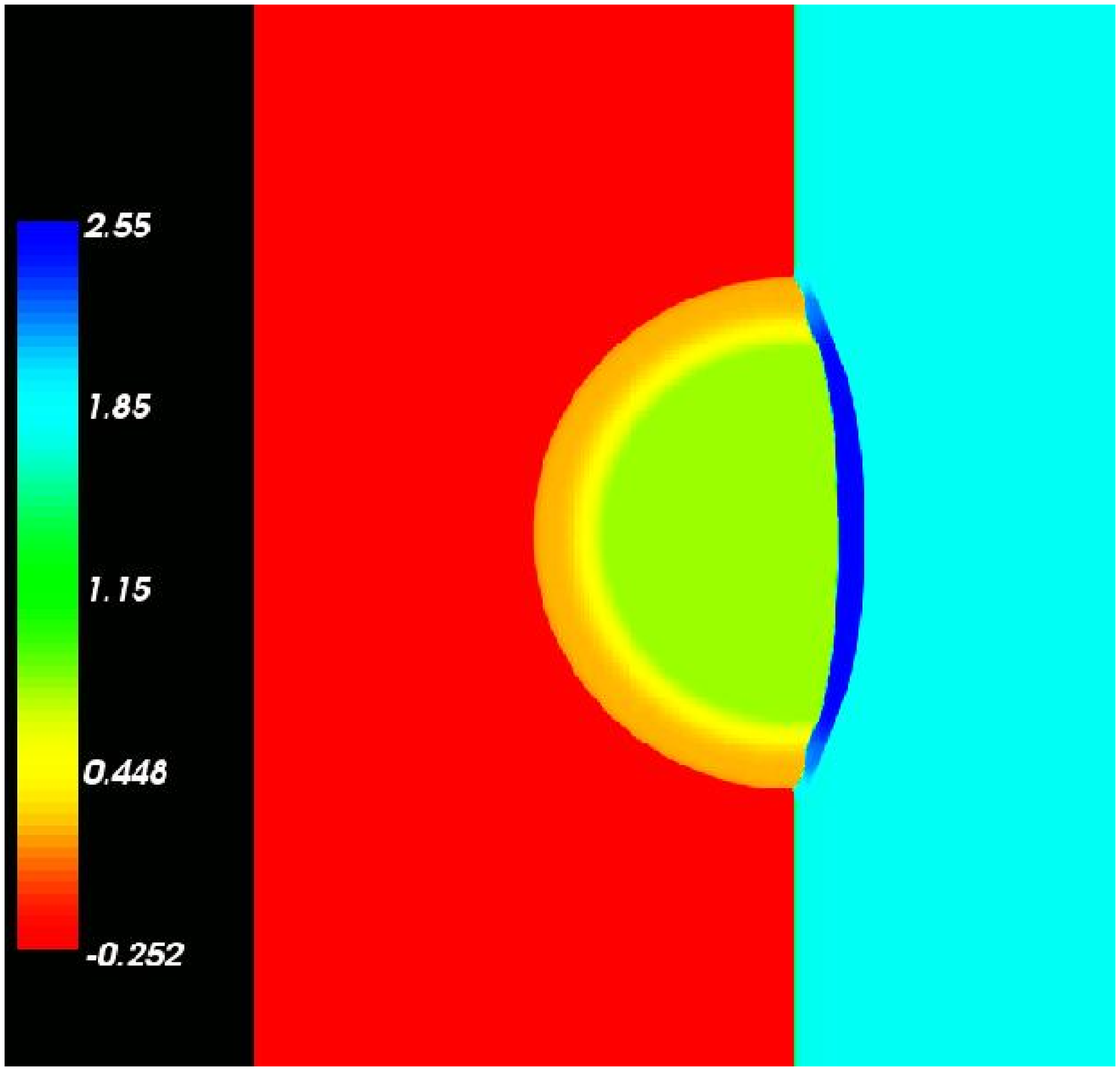}
\caption{Logarithmic pressure maps from the shock-cloud interaction run. 
The shock Mach number is 10 and the
cloud overdensity is 10. The left panel shows the `isothermal' case
with $K=10^8$ and the right panel shows the case in which 
$K^{-1}\simeq \Delta x/v_{shock}$, 
i.e. the relaxation time is comparable to the 
shock cell crossing time. These calculations were 
performed with an AMR code which employed 
a base grid of 256$\times$256 zones and two additional 
levels of refinement with refinement ratio 2.
\label{shockcloud:fig}}
\end{center}
\end{figure}
We use a
base grid of 256$\times$256 zones and allow for two additional levels
of refinement with refinement ratio 2 in regions where the undivided,
relative density gradients, $\Delta \rho/\rho$, exceed 20\%. 

We begin assuming that 
a stiff relaxation term of the form in Eq.~(\ref{tcc:eq}) acts
on the flow internal energy and we consider both the case of a
exceedingly large transfer coefficient, $K=10^8$, as well as the case
in which the relaxation time is comparable to the shock cell
crossing time.
This requires, roughly, that 
\begin{equation}
K^{-1}\simeq \Delta x/u_{shock} 
\end{equation}
where $\Delta x$ is the mesh size and $u_{shock}$ is the shock speed.
Note that the stiffness is sufficiently large that refining by a factor 
of 2 or 4 does not make the problem significantly less stiff.
At simulation start the temperature is constant throughout the domain, so
that the cloud is in thermal equilibrium but it has higher pressure
than its surroundings.  As a result, it expands sonically into the
background.  The shock propagates from the right to the left along the
x-axis and as it runs into the cloud it crushes it.  In
Fig.~\ref{shockcloud:fig} we plot the logarithmic pressure map as the
shock is roughly half-way through the cloud. The high pressure
postshock region is clearly thinner in the case of the larger value of
$K$ and the shock has also propagated slightly further down the axis.
In both cases, however, the result is sound and shows no sign of numerical artifact
both in the presence of strong shock and large gradients, and
independently of the magnitude of the transfer coefficient.
\begin{figure} 
\begin{center}
\includegraphics[width=0.50\textwidth,  scale=1.0]{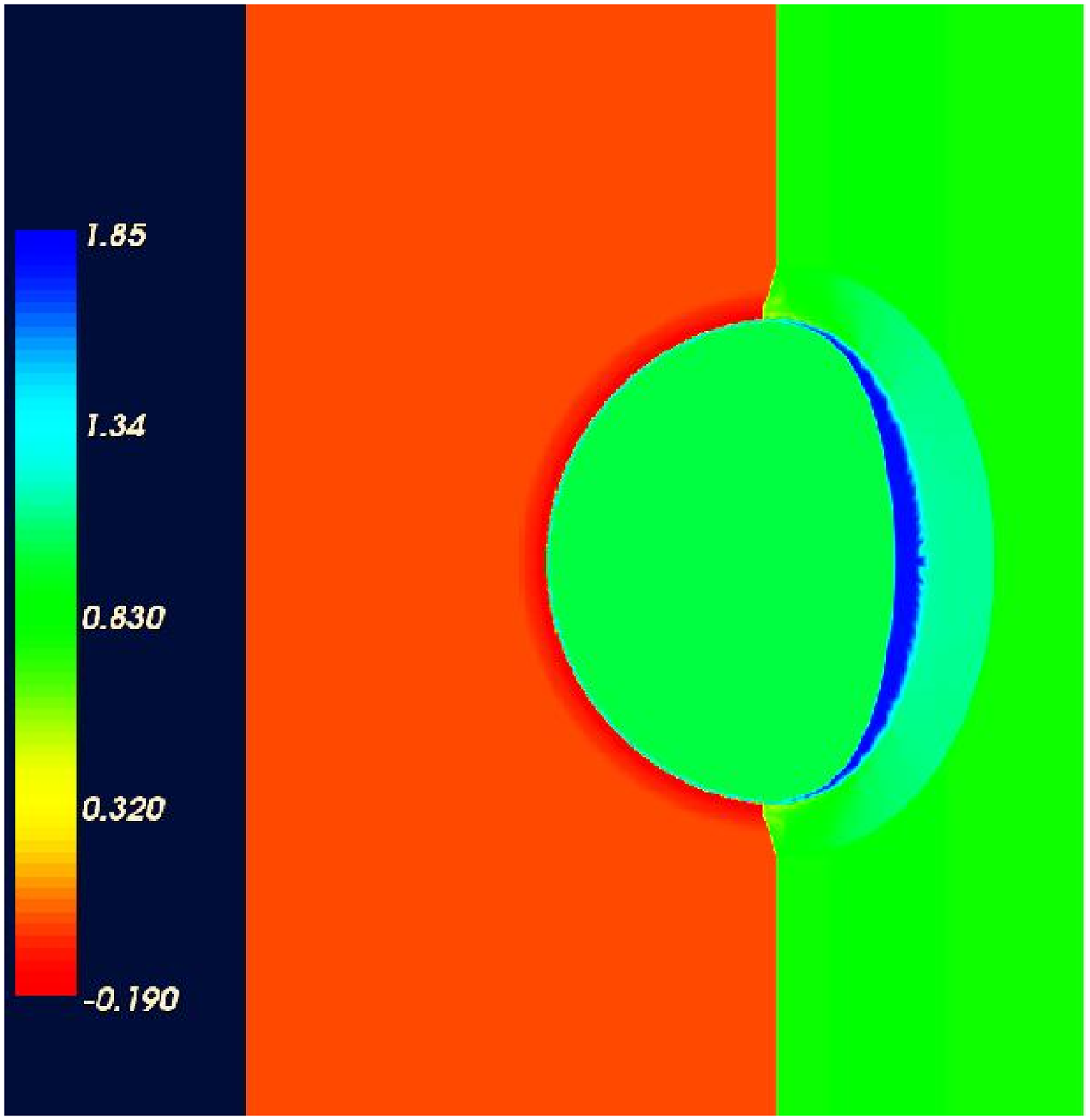}\includegraphics[width=0.50\textwidth, scale=1.0]{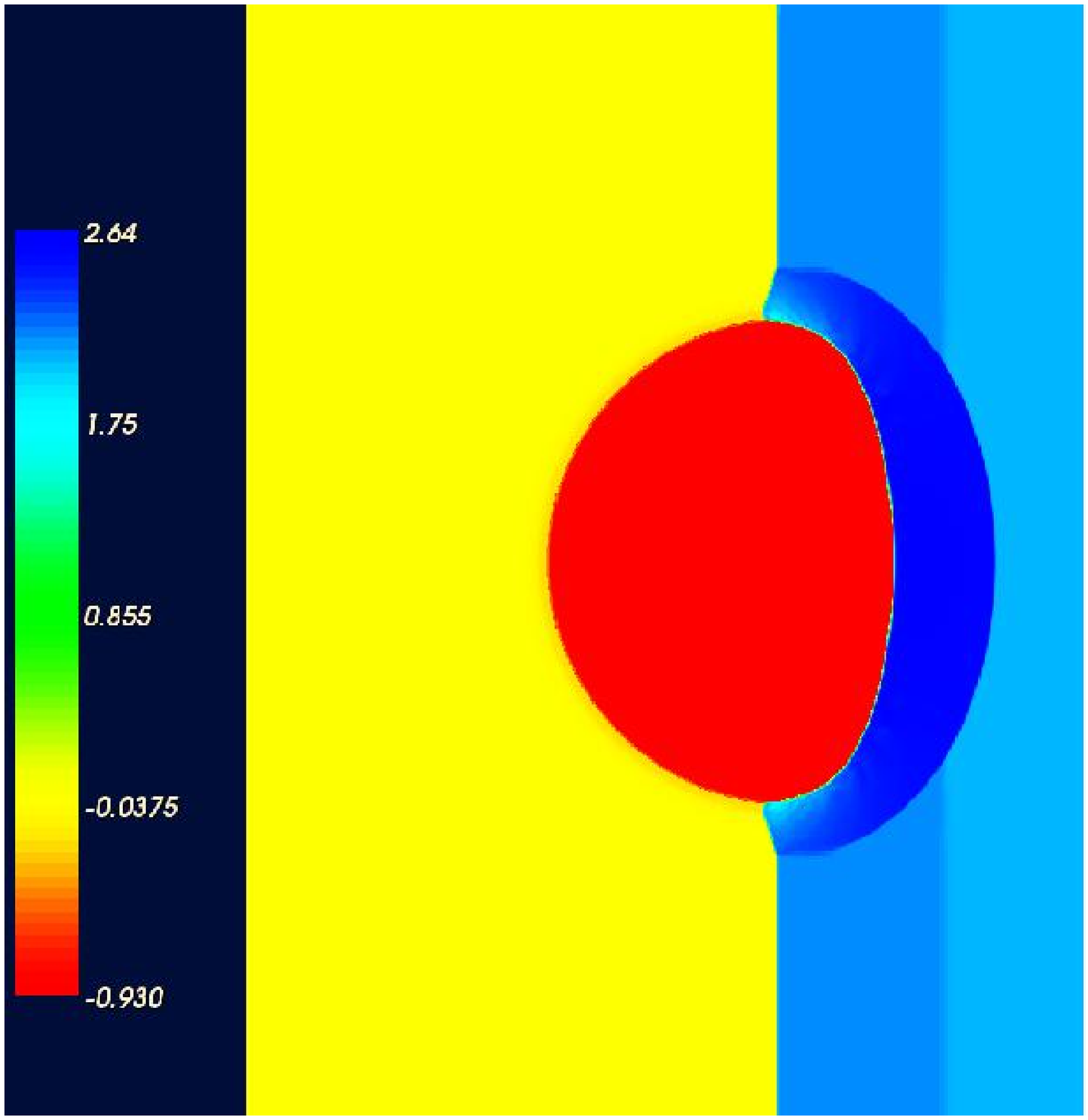}\\
\includegraphics[width=0.50\textwidth, scale=1.0]{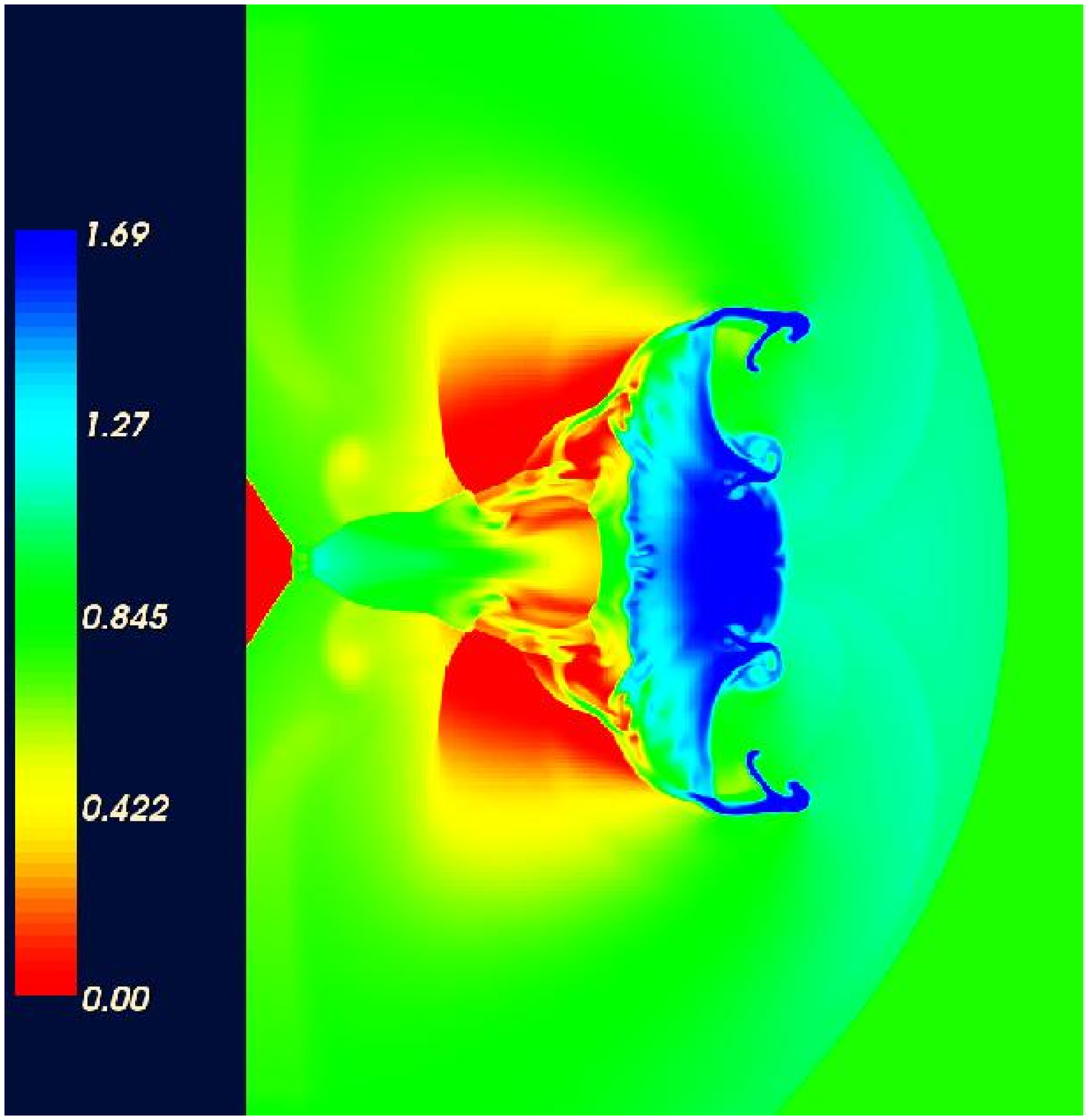}\includegraphics[width=0.50\textwidth, scale=1.0]{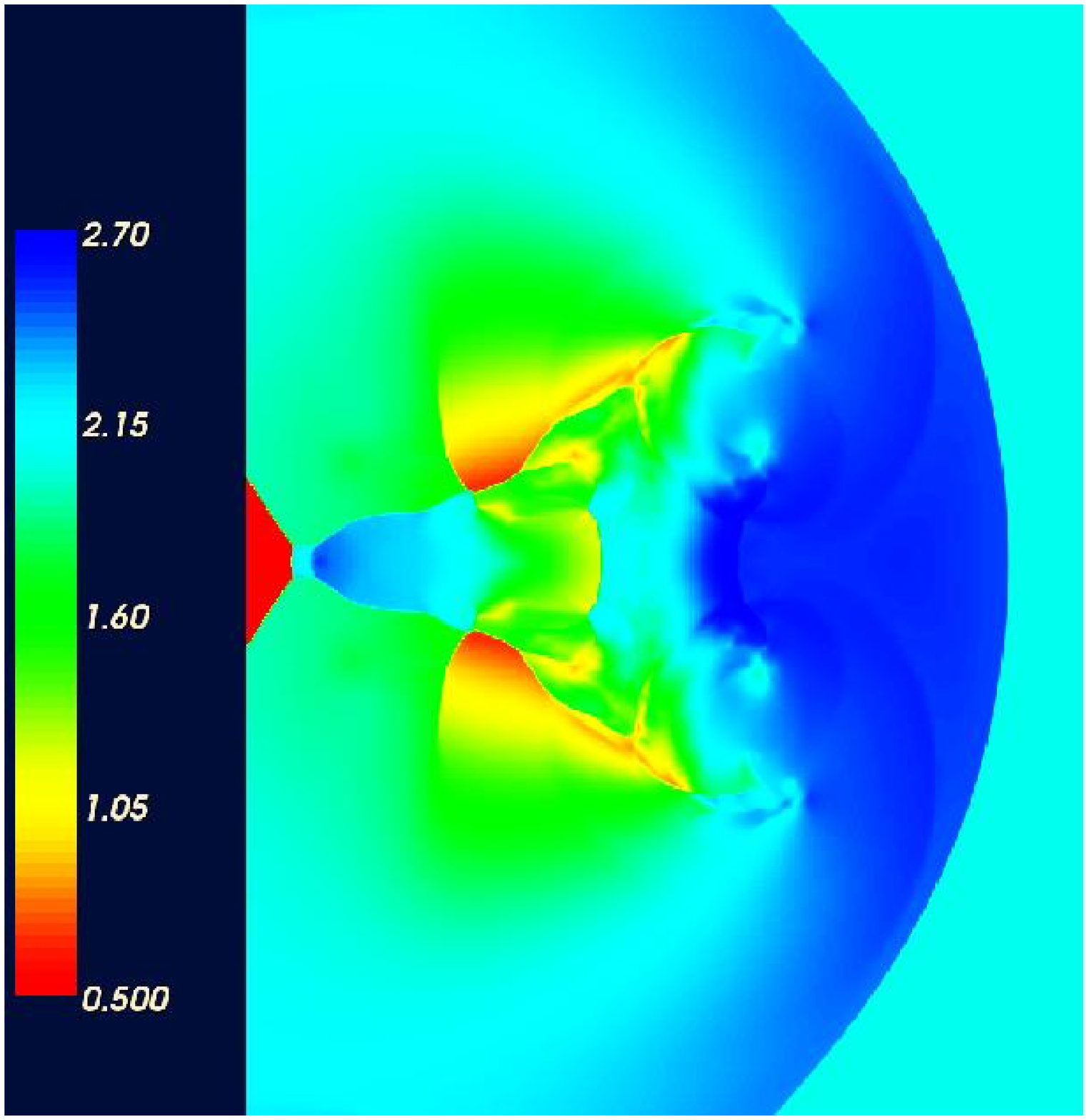}
\caption{Logarithmic density (left) and pressure (right) maps
from the shock-cloud interaction run for a density dependent source term
at $t=0.018$ (top) and $t=0.07$ (bottom) time units.
As before, the shock Mach number is 10, the cloud overdensity is 10
and the calculation was performed with an AMR code employing 
a base grid of 256$\times$256 zones and two additional 
levels of refinement with refinement ratio 2.
\label{shockcloud2:fig}}
\end{center}
\end{figure}

As a final test, we consider the same shock cloud interaction problem
as described above but with a source function appropriate for a
mixture of hydrogen (76~\%) and helium (24~\%) illuminated by a
uniform ionizing background radiation field. The cooling
part of the source function
is proportional to the density and the equilibrium temperature,
of order $10^4$ K, depends slightly on the density. 
This function has a very strong temperature gradient about the 
equilibrium value, behaving analogously to the stiff source terms
used in the previous sections were accuracy and convergence studies
were carried out.

We set the background
gas temperature to $10^6$ K and its number density to 0.1 cm$^{-3}$.
The gas is collisionally ionized, its sound speed is of order
$1.2\times 10^7$ cm s$^{-1}$ and it has a cooling time $\tau_{cool} =
P/(\gamma-1) \rho \Lambda \simeq 1.4\times 10^7$ yr.  With a box
size $L=1.7$ kpc = $5.2 \times 10^{21}$ cm, the latter is much longer
than the CFL time, $\tau_{CFL} \leq \Delta x / u_{shock} \simeq
5.4\times 10^3$ yr.  The cloud of overdense gas is in pressure
equilibrium with a density contrast $\chi=10$ so that its temperature
is $10^5 $ K.  When unperturbed, the cloud's gas cooling time is about
$1.4\times 10^4$ yr $\simeq 2.6\times\tau_{CFL}$. A background radiation
field, producing about $\Gamma\sim 2.4\times 10^{-12}$ s$^{-1}$
ionizations of neutral atoms of hydrogen and helium keeps the 
cloud's temperature at the equilibrium value of $\sim 1.5 10^4$ K.
However, the cloud's pressure quickly falls below the background value
and, as a minor effect, the cloud slowly contracts.

Fig.~\ref{shockcloud2:fig}, shows a snapshot of the density (left) and
the pressure (right) during the initial (top) and final stages
(bottom) of the simulation.  The reverse shock is non-radiative, thus
extending further ahead of the cloud than in the previous cases in
Fig.~\ref{shockcloud:fig}.  Inside the cloud strong
radiative losses prevent the full temperature rise in the postshock
region and produce a density jump substantially larger than in the
corresponding adiabatic case.
The bottom panel shows the later stages of the cloud evolution, when
Rayleigh-Taylor instability with scales comparable to the cloud size
have developed and are shredding the cloud.
As in the previous case, in
which the source term is described by a relaxation law, the code
appears to produce reliable numerical results, without numerical
artifact despite the presence of strong shock and large gradients.

\section{Conclusions} \label{concl:se}

We have presented a second order accurate semi-implicit
predictor-corrector scheme to treat stiff source terms within the
framework of higher order Godunov's methods.  Our treatment of the
predictor step for computing the hyperbolic fluxes, is based on the
derivation of a local effective dynamics using Duhamel's formula.
This leads to a conventional second-order Godunov method when the
system relaxation time is larger than the time step and to a
second-order Godunov method for the isothermal equations in the limit
of a stiff source term.  Finally, we obtain a semi implicit corrector
using a one-step second-order accurate deferred corrections method as
suggested in \cite{dugrro00,minion03}.
 
Our tests indicate that the proposed method is stable, robust and
its second order accuracy preserved across a variety of stiffness
conditions.  We have also discussed the case of a general source term
which depends both on $e$ and $\rho$ and shown that the method is
applicable provided that the flow is thermally stable or the
non-stiff part of the source term is resolved in time.

The additional cost involved in the formulation of our scheme is 
minimal; all it requires is an estimate of the term $\Lambda_e$ which in
a purely relaxation case is trivial and for a more complicated
source term (such as the case of radiative losses) is still minor
compared to the estimate of the source term itself. In our implementation
the factor $\alpha(\gamma-1)+1$ is stored as an 
additional primitive variable and used as
polytropic index in the characteristic analysis in stead of
$\gamma$.

\vskip 1truecm
\leftline{\bf Acknowledgment}
FM is grateful to the Lawrence Berkeley National Laboratory
for its hospitality and 
acknowledges support by the Swiss Institute of Technology through
a Zwicky Prize Fellowship. PC was supported by the Mathematical,
Information, and Computing Sciences Division of the United States
Department of Energy Office of Science under contract number
DE-AC02-05CH11231.

\bibliographystyle{plain}
\bibliography{../biblio/books,../biblio/codes,../biblio/papers}

\begin{thebibliography}{10}

\bibitem{bergercolella89}
M.~J. Berger and P.~Colella.
\newblock Local adaptive mesh refinement for shock hydrodynamics.
\newblock {\em \jcp}, 82:64--84, 1989.

\bibitem{caflisch97}
R.~E. Caflisch, S.~Jin, and G.~Russo.
\newblock Uniformly accurate schemes for hyperbolic systems with relaxation.
\newblock {\em SIAM J. Sci. Comput.}, 34(1):246--281, 1997.

\bibitem{chleli94}
G.~Q. {Chen}, C.~{Levermore}, and T.~{Liu}.
\newblock {Hyperbolic conservation laws with stiff relaxation terms and
  entropy}.
\newblock {\em Comm. Pure Appl. Math.}, 47:787--830, 1994.

\bibitem{colella90}
P.~Colella.
\newblock Multidimensional upwind methods for hyperbolic conservation laws.
\newblock {\em \jcp}, 82:64--84, 1989.

\bibitem{dugrro00}
A.~{Dutt}, L.~{Greengard}, and V.~{Rokhlin}.
\newblock {Spectral deferred correction methods for ordinary differential
  equations}.
\newblock {\em BIT}, 40:241--266, 2000.

\bibitem{field65}
G.~B. Field.
\newblock Thermal instability.
\newblock {\em \apj}, 142:531, 1965.

\bibitem{jin95}
S.~Jin.
\newblock Runge-kutta methods for hyperbolic conservation laws with stiff
  relaxation terms.
\newblock {\em J. Comput. Phys.}, 122:51--67, 1995.

\bibitem{jinlevermore96}
S.~Jin and D.~Levermore.
\newblock Numerical schemes for hyperbolic conservation laws with stiff
  relaxation terms.
\newblock {\em J. Comput. Phys.}, 126:449--467, 1996.

\bibitem{jinpareschitoscani98}
S.~Jin, L.~Pareschi, and G.~Toscani.
\newblock Diffusive relaxation schemes for multiscale discrete-velocity kinetic
  equations.
\newblock {\em SIAM J. Sci. Comput.}, 35(6):2406--2439, 1998.

\bibitem{mico06a}
F.~Miniati and P.~Colella.
\newblock Block structured adaptive mesh and time refinement for hybrid,
  hyperbolic + n-body systems.
\newblock {\em \jcp}, 2006.
\newblock submitted.

\bibitem{minetal00}
F.~Miniati, D.~Ryu, H.~Kang, T.~W. Jones, R.~Cen, and J.~Ostriker.
\newblock Properties of cosmic shock waves in large-scale structure formation.
\newblock {\em \apj}, 542:608--621, 2000.

\bibitem{minion03}
M.~Minion.
\newblock Semi-implicit spectral deferred correction methods for ordinary
  differential equations.
\newblock {\em Comm. Math. Sci.}, 1:471--500, 2003.

\bibitem{pareschirusso05}
L.~Pareschi and G.~Russo.
\newblock Implicit---explicit runge---kutta schemes and applications to
  hyperbolic systems with relaxation.
\newblock {\em SIAM J. Sci. Comput.}, 25(1):129--155, 2005.

\bibitem{pember93}
R.~B. Pember.
\newblock Numerical methods for hyperbolic conservation laws with stiff
  relaxation ii: higher-order godunov methods.
\newblock {\em SIAM J. Sci. Comput.}, 14(4):824--859, 1993.

\bibitem{roehit01}
P.~L. {Roe} and A.~F. {Hittinger}.
\newblock {Toward Godunov-type methods for hyperbolic conservation laws with
  stiff relaxation}.
\newblock In E.~F. Toro, editor, {\em Godunov Methods: Theory and
  Applications}, pages 725--744, New York, 2001. Kluwer Academic/Plenum
  Publishers.

\bibitem{saltzman94}
J.~{Saltzman}.
\newblock {An Unsplit 3D Upwind Method for Hyperbolic Conservation Laws}.
\newblock {\em Journal of Computational Physics}, 115:153--168, November 1994.

\bibitem{tcm05}
D.~{Trebotich}, P.~{Colella}, and G.~H. {Miller}.
\newblock {A stable and convergent scheme for viscoelastic flow in contraction
  channels}.
\newblock {\em Journal of Computational Physics}, 205:315--342, May 2005.

\bibitem{vincentiKruger65}
W.~G. Vincenti and C.~H. Kruger.
\newblock {\em Introduction to Physical Gas Dynamics}.
\newblock John Wiley, New York, 1965.

\bibitem{withman74}
G.~B. Withman.
\newblock {\em Linear and Non-Linear Waves}.
\newblock Wiley-Interscience, New York, 1974.

\bibitem{woodwardcolella84}
P.~R. Woodward and P.~Colella.
\newblock Numerical simulations of two-dimensional fluid flow with strong
  shocks.
\newblock {\em \jcp}, 54:115--173, 1984.

\end{thebibliography}

\end{document}